\documentclass[sigconf]{acmart}

\usepackage{booktabs} 
\usepackage{algorithm}
\usepackage{subcaption}
\usepackage[noend]{algpseudocode}
\usepackage{balance}
\usepackage{multirow}
\usepackage{colortbl}

\renewcommand\footnotetextcopyrightpermission[1]{} 
\makeatletter
\def\BState{\State\hskip-\ALG@thistlm}
\renewcommand\@formatdoi[1]{\ignorespaces}
\makeatother

\setcopyright{rightsretained}

\begin{document}
\title[Save Our Spectrum]{Save Our Spectrum: Contact-Free Human Sensing\\Using Single Carrier Radio}


\author{Alemayehu Solomon Abrar}
\affiliation{%
  \institution{University of Utah}
  \city{Salt Lake City} 
  \state{UT} 
}
\email{aleksol.abrar@utah.edu}

\author{Anh Luong}
\affiliation{%
  \institution{Carnegie Mellon University}
  \city{Pittsburgh} 
  \state{PA} 
}
\email{anhluong@cmu.edu}

\author{Peter Hillyard}
\affiliation{%
  \institution{Xandem}
  \city{Salt Lake City} 
  \state{UT} 
}
\email{peter@xandem.com}

\author{Neal Patwari}
\affiliation{%
  \institution{University of Utah and Xandem}
  \city{Salt Lake City} 
  \state{UT} 
}
\email{npatwari@ece.utah.edu}


\begin{abstract}
Recent research has demonstrated new capabilities in radio frequency (RF) sensing that apply to health care, smart home, and security applications. However, previous work in RF sensing requires heavy utilization of the radio spectrum, for example, transmitting thousands of WiFi packets per second. In this paper, we present a device-free human sensing system based on received signal strength (RSS) measurements from a low-cost single carrier narrowband radio transceiver. We test and validate its performance in three different applications: real-time heart rate monitoring, gesture recognition, and human speed estimation. The challenges in these applications stem from the very low signal-to-noise ratio and the use of a single-dimensional measurement of the channel.  We apply a combination of linear and non-linear filtering, and time-frequency analysis, and develop new estimators to address the challenges in the particular applications.
Our experimental results indicate that RF sensing based on single-carrier magnitude measurements performs nearly as well as the state-of-the-art while utilizing three orders of magnitude less bandwidth.
\end{abstract}

\keywords{Device-free, gesture recognition, heart rate estimation, received signal strength, RF sensing, speed estimation.}

\maketitle

\section{Introduction}

Radio frequency (RF) based, device-free human sensing is the use of radio signals to detect human vital signs and classify human activities without requiring a user to carry or wear any sensors. Such sensing has been shown to be useful in non-invasive human vital sign monitoring \cite{adib2015smart, liu2015tracking, patwari2014breathfinding}, 
device-free localization \cite{kaltiokallio2012follow, yang2013rssi, adib20143d}, fall detection and elder care \cite{mager2013fall,kaltiokallio2012follow}, and gesture and activity recognition in a smart home \cite{wang2015understanding, pu2013whole}. 
Due to the ubiquity and low cost of radio transceivers, and the surprising success and accuracy reported for systems using these applications, we can foresee many future commercial systems being used in daily life.  RF sensing has advantages with respect to privacy and coverage compared to vision-based sensors.  The use of RF signals for sensing applications, however, requires utilization of bandwidth which is currently dominated by wireless communication. RF sensing must compete for use of the spectrum, and a system that utilizes less bandwidth for these applications becomes extremely important. 

A variety of RF-based sensors have been proposed, including 
radar-based systems \cite{ kilic2014device, lazaro2010analysis, adib20143d, adib2015smart, molchanov2015short, ren2015noncontact, ram2008doppler}, WiFi-based systems \cite{liu2015tracking, wang2015understanding, pu2013whole}, and systems based on received signal strength (RSS) measurements \cite{kaltiokallio2014non, kaltiokallio2012follow, sigg2014rf, patwari2014breathfinding, mager2013fall}. Most of these systems require heavy utilization of the spectrum. Radar-based solutions typically require a bandwidth on the order of GHz to operate. For instance, the frequency-modulated continuous-wave (FMCW) system in \cite{adib2015smart} provides accurate breathing and heart rate estimates by utilizing $1.79$ GHz of bandwidth. Spectrum utilization of WiFi-based systems is also inefficient, as  WiFi systems such as 802.11 a/g/n are based on utilization wideband 20 or 40 MHz channels \cite{pu2013whole,liu2015tracking,wang2015understanding}. For example, Wang \emph{et al.} present CARM \cite{wang2015understanding}, a WiFi-based activity recognition system which requires transmission of 2500 WiFi packets every second.  This represents 62.5\% utilization of a 20 MHz channel.
Such channel utilization makes WiFi-based RF sensing systems difficult to operate in a typical indoor environment simultaneously with many existing WiFi communications systems.
RSS-based systems have also been used in breathing monitoring \cite{kaltiokallio2014non, patwari2014breathfinding, luong2016rss} and localization \cite{kaltiokallio2012follow, wilson2011see}. However, despite their efficient spectrum utilization compared to other RF sensing systems, most RSS-based systems process coarse-grained RSS data, which  are unable to capture mm-level motions such as skin vibrations due to heartbeat. 

RF-based sensing systems are to be deployed in our mobile devices for gesture recognition, in our homes and businesses for activity monitoring, in localization and security, and by our beds for sleep breathing monitoring.  Simultaneously, the same bands are seeing rapidly increasing utilization for data communications \cite{cisco2015forcast}. 
The utilization of significant portions of our unlicensed bands for RF sensing, for example, using 20 MHz for a single sensor, will be very challenging to realize. 

The purpose of this paper is to show that a narrowband RF sensing system, which uses three orders of magnitude less spectrum, can provide similar performance for multiple applications.  Demonstrating this in a low-cost device has the potential to enable widespread consumer use of RF sensing without a significant impact to existing wireless communications systems.
We propose a system containing a single pair of devices equipped with commercial-off-the-shelf (COTS) narrowband radio transceivers.
Each device uses a TI CC1200 radio attached to a Beaglebone processor to obtain, using code downloaded from \cite{cc1200code}, a high resolution measure of RSS.
We show the capabilities of this system in real-time pulse rate estimation, gesture recognition and speed estimation while utilizing only $11.26$ kHz of the spectrum. 

Our system determines the heart rate of a user by applying a fast Fourier transform (FFT) to the RSS data, and finding the peak frequency from a modified power spectral density (PSD) measure in the cardiac frequency band. Gesture recognition is performed first by applying variance-based segmentation to detect gesture regions in the RSS data followed by time-frequency analysis using the discrete wavelet transform (DWT) to extract features. We show the performance of our system under three different classifiers. We also present a method to estimate human walking speed from the spectral components of single-channel RSS measurements.   

\textbf{Challenges and Solutions.} Extracting RF sensing information from single-channel magnitude measurements is challenging for multiple reasons. 
First, the RSS data is highly corrupted by both Gaussian and non-Gaussian noise. Without any processing, it is often difficult to extract weak signals from RSS measurements. To denoise the RSS data, we employ a combination of Hampel and Butterworth filters. A Hampel filter is used to remove outliers \cite{liu2015tracking} while a Butterworth bandpass filter is applied to cancel out-of-band noise, respiration harmonics and other low-frequency interfering signals.
Second, changes in a link's signal strength caused by the heartbeat-induced vibration of a person's skin is very weak, and its energy is distributed along multiple harmonics of its fundamental frequency. For these reasons, pulse rate has not, to our knowledge, ever been reported to be estimated from RSS measurements. To enhance heartbeat detection, we use directional antennas to focus more RF power on the person's body, and we introduce a method to combine the harmonics in the magnitude spectrum to improve estimation performance.
Third, single-channel measurements by their nature do not provide high dimensionality such as obtained in state-of-the-art wideband approaches. Different people may perform the same gesture differently, and high-dimensional measurements allow gesture classifiers to perform well despite such differences. We use the high resolution of our measurements, rather than high dimension, and apply appropriate  time-frequency analysis for feature selection and show performance similar to those obtained using other RF sensing systems. 
Finally, extracting human speed from RSS spectral components is also challenging because, during walking, different body parts, e.g., legs, torso and hands, have different instantaneous speeds, thus causing different Doppler effects. By studying the spectrogram of the RSS waveform when a person crosses a link, we develop a method to accurately estimate the average body speed from the average frequency of the RSS data. 


Adopting these methods into our system enables accurate heart rate detection, gesture recognition, and human speed estimation. Heartbeat tests on multiple subjects lying down on a cot result in an average error of 1.5 beats per minute (bpm) across three subjects.  Our gesture recognition algorithm provides an average classification accuracy of 85\% for eight distinct gestures. With our system, human walking speed can also be estimated with an error of only 5.1~cm/s. These results show that accurate RF sensing can be achieved without the use of excessive bandwidth or complex hardware.

In a nutshell, our contributions are as follows:
\begin{itemize}
    \item We introduce an RSS-based heart rate monitoring system. To the best of our knowledge, our system is the first heart rate monitoring system that uses RSS measurements, and it utilizes the lowest bandwidth of any reported RF-based heart rate monitor.
    
    \item We develop a human gesture recognition system that classifies eight different gestures using single-channel RSS measurements.

    \item We propose a novel method to estimate human walking speed from the spectral components of single-channel RSS measurements.

    \item Critically, our work shows that RF sensing can be performed successfully with orders of magnitude less bandwidth than reported in related work, a result which will be critical if RF sensing is to become ubiquitous.

\end{itemize}

\textbf{Roadmap.} In Section \ref{sec:related}, we provide a summary of related work. Section \ref{sec:overview} provides an overview of our RF sensing system. We introduce our RSS-based heart rate estimation in Section \ref{sec:heart}, followed by RSS-based gesture recognition in Section \ref{sec:gesture}. Section \ref{sec:speed} presents human walking speed estimation. We discuss the limitations of our system and future work in Section \ref{sec:discussion} and then conclude the paper.

\section{Related Work}
\label{sec:related}

Most recent applications of device-free human sensing make use of sensors that detect perturbations of visible light, sound, or radio waves due to certain human activities. Visible light-based approaches may employ either cameras \cite{song2015joint,kim2014retrodepth, song2014air} or a combination of light emitting diodes (LEDs) and photo detectors \cite{li14epsilon, kuo2014luxapose, li2015human, zhang2015extending}. Camera-based systems apply computer vision techniques on high-resolution images to track human motion or detect certain activities, while shadow patterns are processed in systems that use LEDs and photo detectors. However, these systems mostly fail due to obstructions by clutters in an indoor environment \cite{li14epsilon}. Additionally, vision-based human sensing often demands high computational power to process the images, and is usually prone to security and privacy threats \cite{sbirlea2013automatic}. On the other hand, sound waves have been exploited in various applications such as indoor localization \cite{tarzia2011indoor, peng2007beepbeep, huang2014shake}, gesture recognition \cite{wang2016device, gupta2012soundwave, yun2015turning}, and breathing monitoring \cite{arlotto2014ultrasonic}. Nevertheless, most of these applications are not device free \cite{tarzia2011indoor, yun2015turning}, or they require the user to be in a very close proximity (in the order of centimetres) to a device \cite{wang2016device, arlotto2014ultrasonic, gupta2012soundwave}. 

Several device-free sensing applications, however, focus on the use of radio frequency (RF) signals, as radio waves can penetrate through walls and cover a large area. RF-based methods employ a set of techniques that extract radio waves that impinge on human subjects from those that interact with static objects. These techniques can be based on received signal strength measurements using standard transceivers, time-of-flight (TOF) or channel response  measurements in radar systems, or channel state information (CSI) or Doppler shift measurements from WiFi devices.

\textbf{Radar-based Systems.} These include impulse radio, ultra-wide band (IR-UWB) \cite{kilic2014device, lazaro2010analysis}, frequency-modulated continuous-wave \cite{van2008feature, adib20143d, adib2015smart, molchanov2015short}, and Doppler \cite{ren2015noncontact, ram2008doppler} radars. Radar systems have been tested in various sensing applications, such as indoor localization \cite{adib20143d,kilic2014device}, vital sign monitoring \cite{adib2015smart, ren2015noncontact, li2013review, lazaro2010analysis}, and gesture recognition \cite{googlesoli, molchanov2015short, kim2009human}. These systems apply TOF or channel impulse response measurements to detect certain human activities. Yet, most of these radar-based systems become impractical in more realistic situations, as they require large bandwidth (in the order of GHz) and high power to operate, and some use expensive, complex hardware, including Universal Software Radio Peripheral (USRP) \cite{adib2015smart,adib20143d}.

\textbf{WiFi-based RF Sensing.} WiFi devices have been utilized in vital sign monitoring \cite{liu2015tracking}, gesture and activity recognition \cite{wang2015understanding, pu2013whole, abdelnasser2015wigest}, and indoor localization \cite{yang2013rssi}. Most of these applications utilize CSI measurements from commercial WiFi devices to detect human motion and activity. These systems, however, utilize large bandwidth and require powerful processing. As an example, Wang \emph{et al.} present DWT based activity recognition system which utilizes a $20$MHz channel. In addition, access to CSI is limited to only few WiFi devices such as Intel 5300 \cite{halperin2011tool}. Other WiFi based systems require the use of expensive hardware as USRP to measure Doppler shifts of multiple pulses in the 20 MHz WiFi channel \cite{pu2013whole}. As a result, the need of tens of megahertz in bandwidth makes WiFi based systems less efficient.

\textbf{Received Signal Strength.} RSS measurements using standard wireless transceivers have been applied in localization and tracking \cite{kaltiokallio2012follow, wilson2011see}, in
breathing monitoring \cite{kaltiokallio2014non, patwari2014breathfinding, luong2016rss}, and in gesture and activity recognition \cite{sigg2014telepathic, sigg2014rf}. In these systems, certain human activities are detected using power measurements taken from standard radio channels such as a 2 MHz zigbee channel or 20 MHz WiFi channel. Most RSS-based systems are generally efficient in terms of bandwidth and power management, but provide only coarse-grained channel information due to quantization. As a result, they are unable to detect mm-level motions, such as heartbeat. Luong \emph{et al.} \cite{luong2016rss} show that small RSS step size enables RF sensing applications, such as breathing monitoring and gesture recognition for a maximum of four gestures. In this paper, we use RSS measurements from single carrier radio to perform heart rate monitoring, gesture recognition for eight gestures, and human speed estimation while utilizing only 11.26 kHz of the spectrum.

\section{System Overview}
\label{sec:overview}

In this section, we describe the hardware and software of the sensing system we use to perform the measurements described in this paper.  We show that the cost of the hardware is low compared to existing systems capable of performing such measurements, and could be further reduced.  We describe the signals transmitted and received, including bandwidth and power, and how the RSS is measured.

\subsection{Platform}
\begin{figure}[tbhp]
  \begin{center}
  \includegraphics[width=0.9\columnwidth]{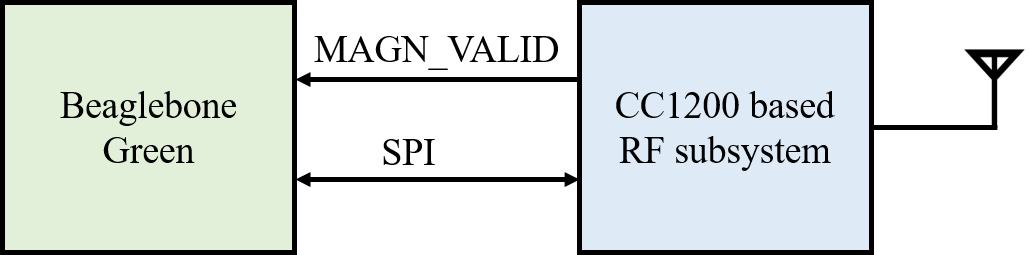}
  \caption{Hardware overview: Each node is composed of a Beaglebone Green and a CC1200 based RF subsystem connected via SPI interface.}
  \label{fig:overview}
  \end{center}
\end{figure}

Our RF sensing hardware is composed of a pair of inexpensive wireless nodes. Each node includes a Beaglebone Green (BBG) platform connected to an RF subsystem via SPI interface as shown in Fig.\ \ref{fig:overview}. 
Our custom RF subsystem is designed as a cape for the BBG, and contains a TI CC1200 sub-GHz narrowband radio transceiver, its matching network, and an SMA-connected antenna.  The CC1200 can operate at multiple center frequencies, including $169$, $434$ MHz and $868$ MHz ISM bands.
The parts for each node cost less than 50 EUR, compared to  more than 1000 EUR for a USRP N200 used in other RF sensing systems \cite{pu2013whole, adib20143d, usrp}.  Further, we note in Section \ref{sec:discussion} that a future streamlined platform design with the same capabilities of our implementation could be built with less than half of the parts cost and significantly lower power consumption as the platform we used here.

\subsection{RSS Measurement}
We reproduce a  high resolution RSS measurement system described in \cite{luong2016rss} using the CC1200 transceiver. In our RF sensing system, we use two devices configured with the code downloaded from \cite{cc1200code}.  One device transmits a CW signal at 434 MHz, while the other captures its receiver's complex baseband samples and sends them to the BBG via SPI.  
The BBG platform includes a programmable real-time processing unit (PRU) sub-system that allows continuous data collection without rate variation. 
The PRU subsystem is programmed to start collecting IQ samples from a CC1200 receiver and write the data onto a shared memory for further processing.

The CC1200 radio transceiver is configured to use a 11.26~kHz channel in the 434~MHz ISM band. The transceiver allows reading the 17-bit magnitude and 10-bit angle of each complex baseband sample received by the CC1200. The PRU subsystem on the BBG is programmed to collect every new sample via SPI bus at a rate of $45$ kHz and store the samples in the shared memory space. Concurrently, the main program running on Linux on the main processor reads data from the shared memory, and adds timestamps. Each RSS estimate, $r(t)$, is computed from 100 samples of the 17-bit magnitude by summing the squared magnitude \cite{cc1200code}.  This process reduces the average sampling rate to $449$ Hz. As the processing algorithm runs in Linux, the sampling period is not purely deterministic. 
However, the RSS sampling rate is more than sufficient for most RF sensing applications, including heart rate monitoring, gesture recognition, and human speed estimation.  

Although not explored in this paper, the sampling rate could enable monitoring of RSS-induced changes up to a few hundred Hz in frequency. Further, while this system uses a CW signal transmission, the CC1200 is also capable of binary FSK data transmission.  This should not affect the results, since FSK is a constant envelope modulation, and would allow the CC1200 to simultaneously be used for data transfer.

\section{Heart Rate Monitoring}
\label{sec:heart}

In this section, we describe the design and performance of a system that estimates the heart rate of a stationary person from measurements of RSS.  Human motion generally changes the RSS on a link between stationary devices, and a heart's beating causes a pulse, i.e., a vibration of arterial blood vessels when blood is pumped through them.  The pulse can be measured as a vibration of the skin, and in fact, the movement of the skin of a person causes slight changes to the radio channel.   Many papers have presented the effects of a person's motion, gesture, and activity on the RSS.  Most closely related, past work has demonstrated that the changes induced by a stationary person's inhaling and exhaling provides sufficient impact on the radio channel to be observed in the RSS on a standard WiFi or Zigbee transceiver \cite{liu2015tracking, li2013review, patwari2014breathfinding, kaltiokallio2014non}.

However, we have not seen a system capable of estimating the heart rate of a person purely from the RSS measured on a stationary link.  We believe this is because of the multiple challenges that need to be overcome to monitor pulse from the very small changes observed in RSS caused by the vibration of a person's skin due to their pulse, including:
\begin{enumerate}
    \item \emph{Low amplitude signal}:  The amplitude of the pulse-induced RSS signal is very small, typically less than  0.01 dB, about an order of magnitude lower than the breathing-induced RSS signal.
    \item \emph{Quantization}: Most commercial transceiver ICs quantize RSS to 1 dB or more.
    \item \emph{Noise}: The noise power in the RSS signal is significantly larger than the pulse signal power.  In addition, the noise is heavy tailed, prone to large impulses.
    \item \emph{Non-sinusoidal waveform}: The movement of the skin due to the pulse more closely resembles a repeating impulse rather than a sinusoid, thus standard spectral analysis is suboptimal.
\end{enumerate}
Our system removes the 1 dB RSS quantization limitation by computing the power from signal complex baseband samples as described in Section \ref{sec:overview}.  In this section, we describe how we apply methods to filter out the noise and interfering signals, and we present a novel spectral domain method to estimate the heart rate despite the non-sinusoidal nature of the signal.

For heart rate estimation, we consider a human subject lying or sitting between a pair of wireless nodes.  Any larger motion of the body would cause very large changes in the RSS that would prevent isolating the pulse-induced signal.  However, humans are stationary at many points during the day and night, and monitoring resting heart rate may have application in home health monitoring, fitness tracking, and sleep monitoring. A person would not need to remember to wear or turn on a heart rate monitor, as non-contact RF sensors could be embedded in the environment and simply monitor heart rate whenever a person was present and stationary.

\begin{figure*}
  \begin{center}
  \begin{subfigure}[t]{0.32\textwidth}
     \includegraphics[width=\textwidth]{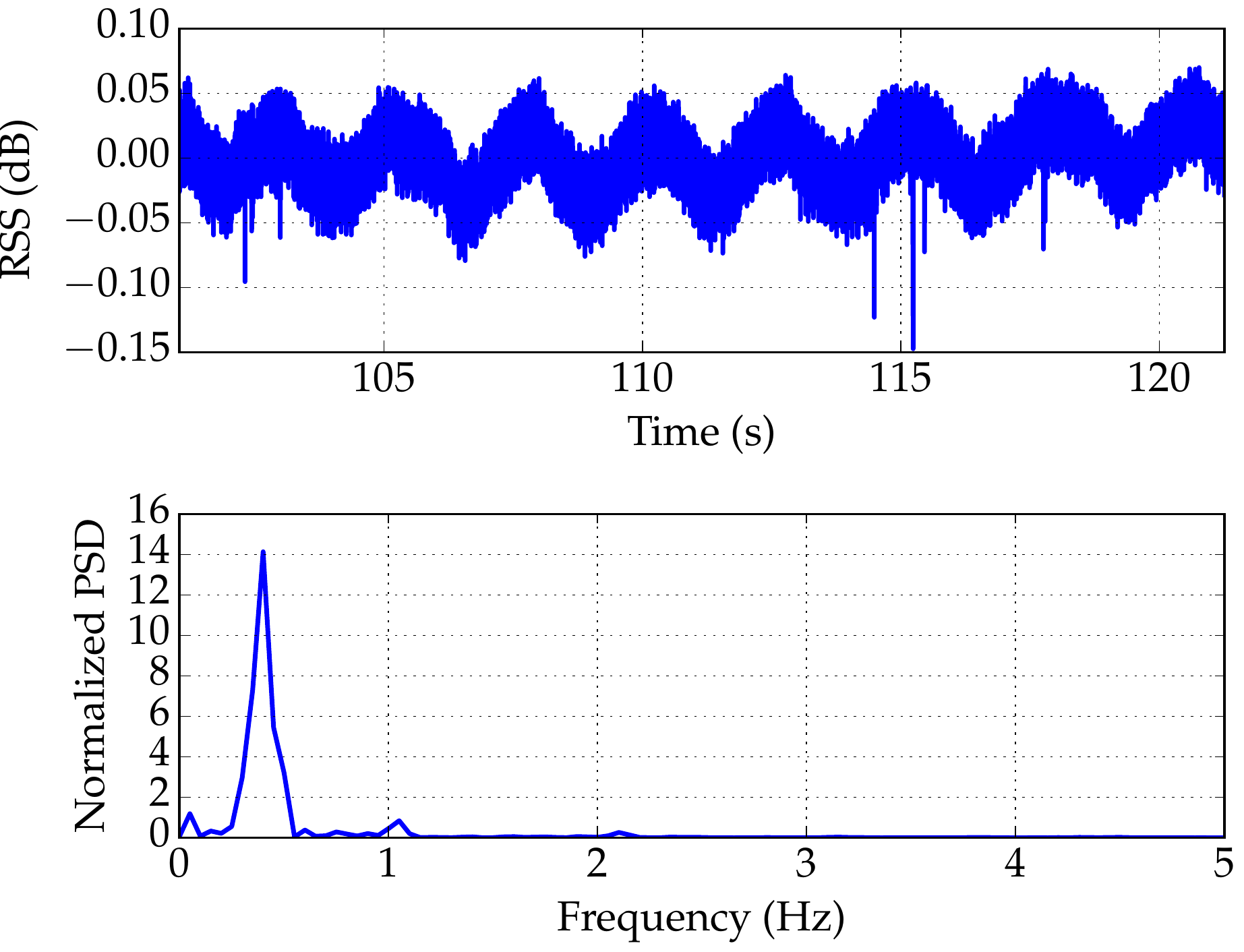}
  \caption{Raw RSS and PSD}
  \label{fig:raw_rss_spec}
  \end{subfigure}
  \hfill
    \begin{subfigure}[t]{0.32\textwidth}
     \includegraphics[width=\textwidth]{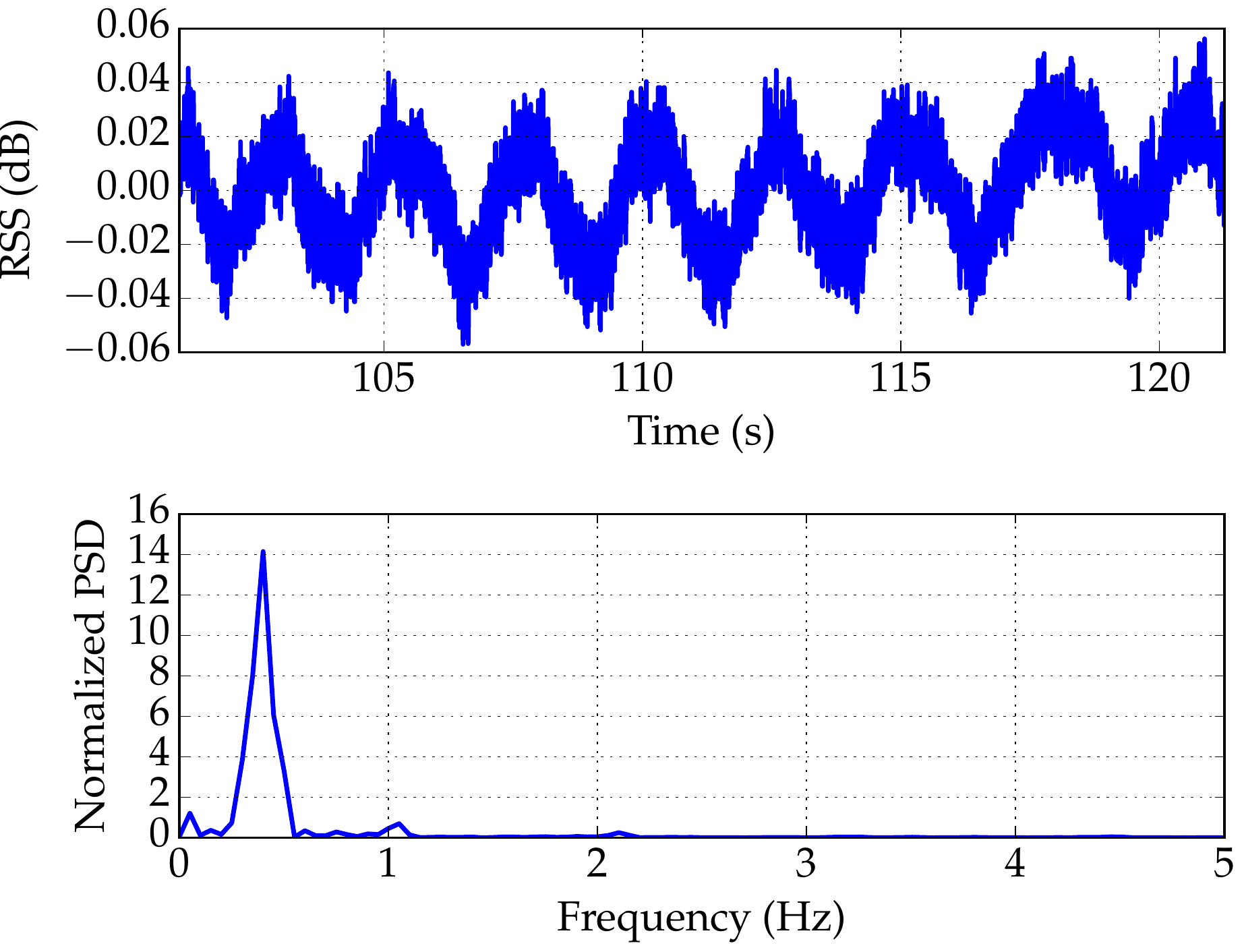}
  \caption{Hampel-filtered RSS and PSD}
  \label{fig:hampel_rss_spec}
  \end{subfigure}
  \hfill
  \begin{subfigure}[t]{0.32\textwidth}
     \includegraphics[width=\textwidth]{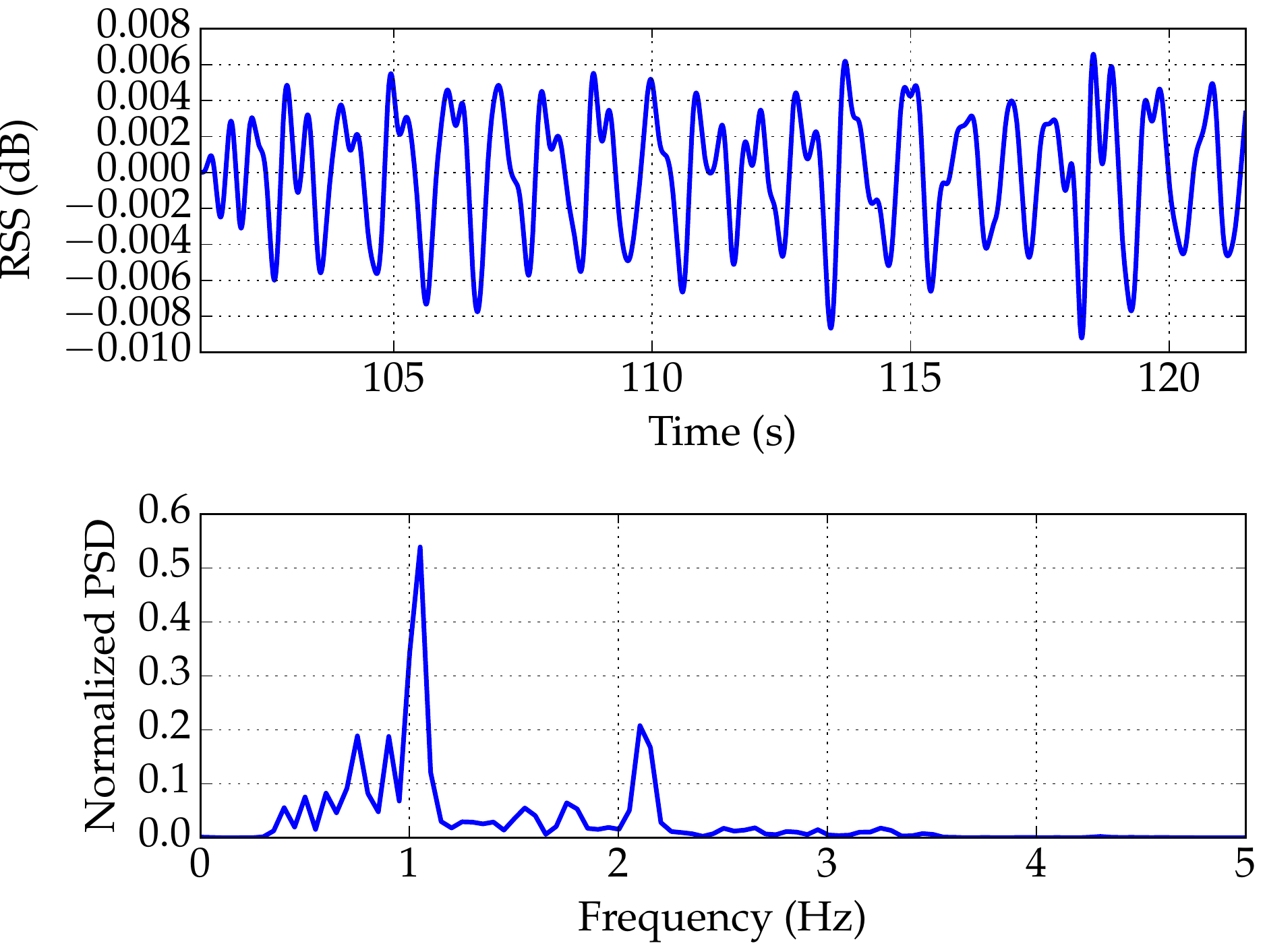}
   \caption{Butterworth-filtered RSS and PSD}
   \label{fig:filtered_rss_spec} 
    
  \end{subfigure}
  
  \label{fig:heart_rss}

\end{center}
  \caption{RSS before and after denoising using Hampel and Butterworth bandpass filters for a person lying down.}
 \end{figure*}

Fig.\ \ref{fig:raw_rss_spec} shows an RSS signal recorded from a seated subject. Breathing patterns are visible in the power measured at the receiver --- we can see about eight cycles with peak-to-peak amplitude of about 0.1 dB over 20 seconds, which corresponds to a breathing rate of 24 breaths per minute. Without any other signal processing, however, it is impossible to identify the pulse signal. In addition to the breathing-induced signal, the RSS measurements are corrupted by noise that appears to be additive and Gaussian, as well as impulsive, heavy-tailed noise that sends the measured RSS significantly lower or higher for one or two samples.

\subsection{Methods}

\subsubsection{RSS Denoising}

We apply denoising techniques to be able to remove the effects of noise and corrupting signals and extract the pulse-induced signal.
To replace outliers due to heavy-tailed noise, we implement a Hampel filter, a generalized median filter with more flexibility in parameter tuning \cite{pearson2016generalized}. 
The window size for the filter should be small enough  to capture fast changing pulses, and long enough to remove sharp outliers. Fig.\ \ref{fig:hampel_rss_spec} shows the output of the Hampel filter for the raw data in Fig.\ \ref{fig:raw_rss_spec}. The Hampel filter can be seen to remove outliers from the raw RSS data, but to keep low-frequency interfering signals, including a strong breathing-induced signal.
Hence, a fourth order Butterworth bandpass filter (BPF) is applied to cancel out additive Gaussian noise, respiration harmonics, and other low-frequency interference. Butterworth filters are chosen for their flat frequency response in the passband. The lower and higher cut-off frequencies of the BPF are set to $0.8$ Hz and $5.0$ Hz, respectively. The PSD of the resulting signal, shown in  Fig.\ \ref{fig:filtered_rss_spec}, is dominated by the harmonics of the pulse signal.

\subsubsection{Heart Rate Estimation}

\begin{figure}[tbhp]
  \begin{center}
  
  \includegraphics[width=0.9\columnwidth]{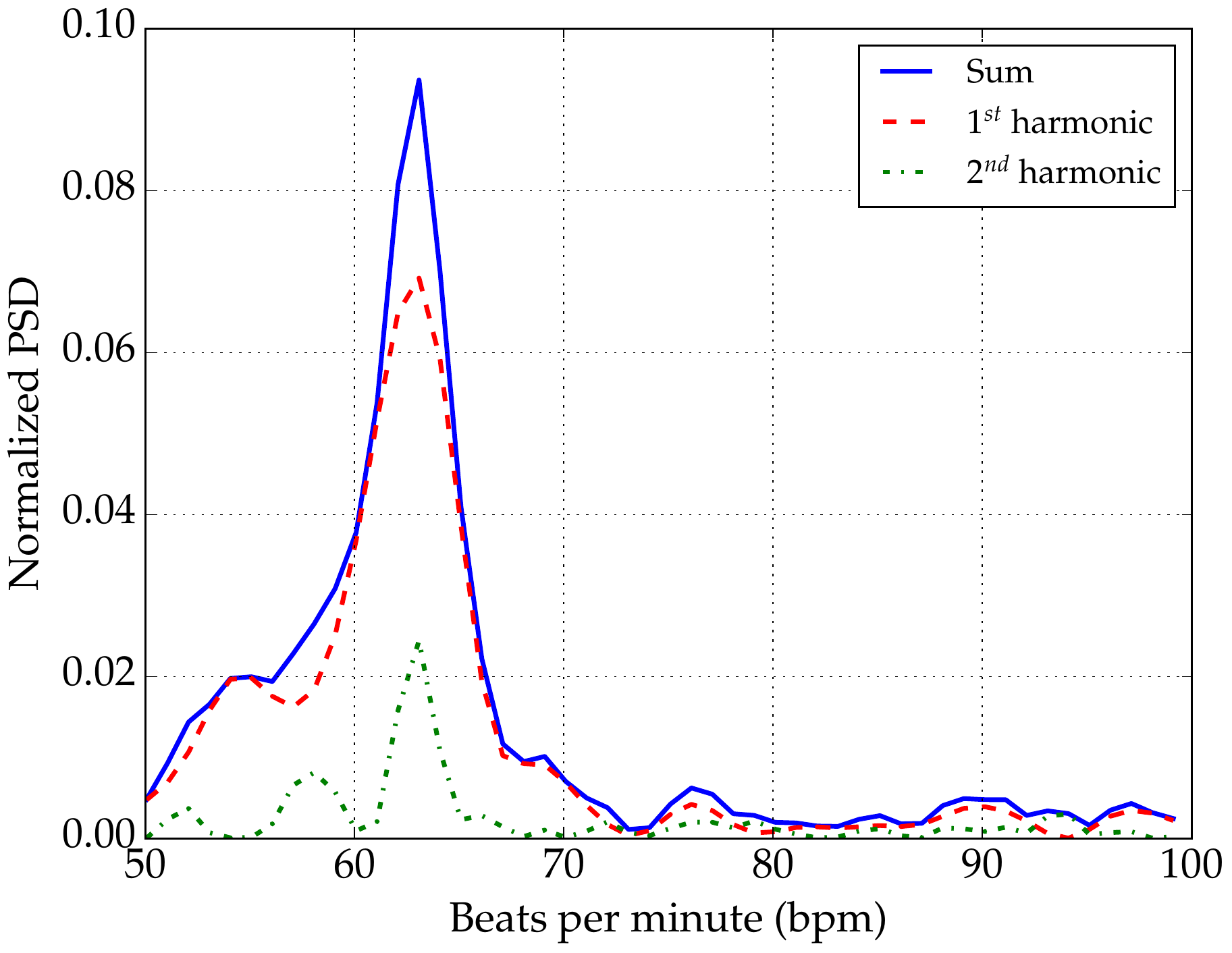}
  
  \caption{Estimate heart rate by superimposing harmonics}
  \label{fig:added_harmonics}
  \end{center}
\end{figure}

\begin{figure}[tbhp]
  \begin{center}
  
  \includegraphics[width=0.9\columnwidth]{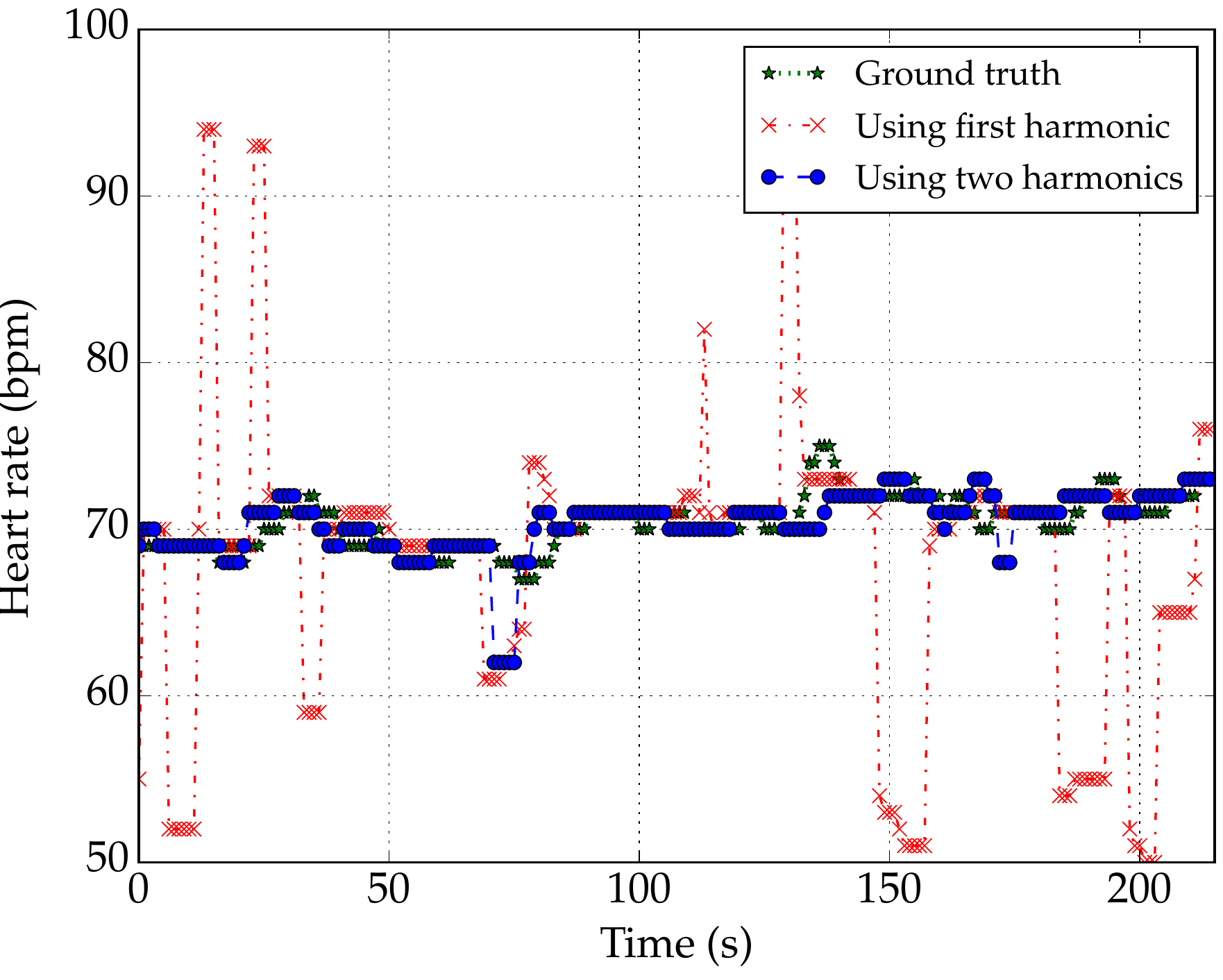}
  
  \caption{Heart rate estimation with and without superimposing harmonics}
  \label{fig:one_harmonic}
  \end{center}
\end{figure}

Once the undesired signals are filtered out from the raw RSS, a resting heart rate can be estimated from the PSD using our estimation algorithm. Since a normal adult human resting heart rate falls in the range between $f_{min} = 0.84$ Hz and $f_{max} = 1.67$ Hz \cite{mason2007electrocardiographic}, the heart rate is estimated by locating the peak value of PSD in the frequency band. However, pulse energy is distributed along harmonics of the fundamental frequency of the pulse signal. Pulse detection is enhanced by superimposing the first two harmonic bands corresponding to cardiac frequency band, as shown in Fig.\ \ref{fig:added_harmonics}. The heart rate is determined by finding the peak frequency in the resulting superimposed PSD. 
As shown in Fig.\ \ref{fig:one_harmonic}, without combining the harmonics, estimation using the first harmonic results in significantly reduced accuracy.

The PSD is calculated from a sliding window of filtered RSS data.
The choice of window size is a trade-off between the ability to track heart rate variability and less dropout as result of small signal-to-noise ratio (SNR). Higher window size increases latency and can result in high error, since it does not allow tracking of the changing heart rate, while very low window size can result in inaccurate heart-rate estimates because of random dropouts, that is, periods when the pulse-induced SNR becomes very small for digitization \cite{li2013review}.

 \begin{figure}[tbhp]
  \begin{center}
  
  \includegraphics[width=0.9\columnwidth]{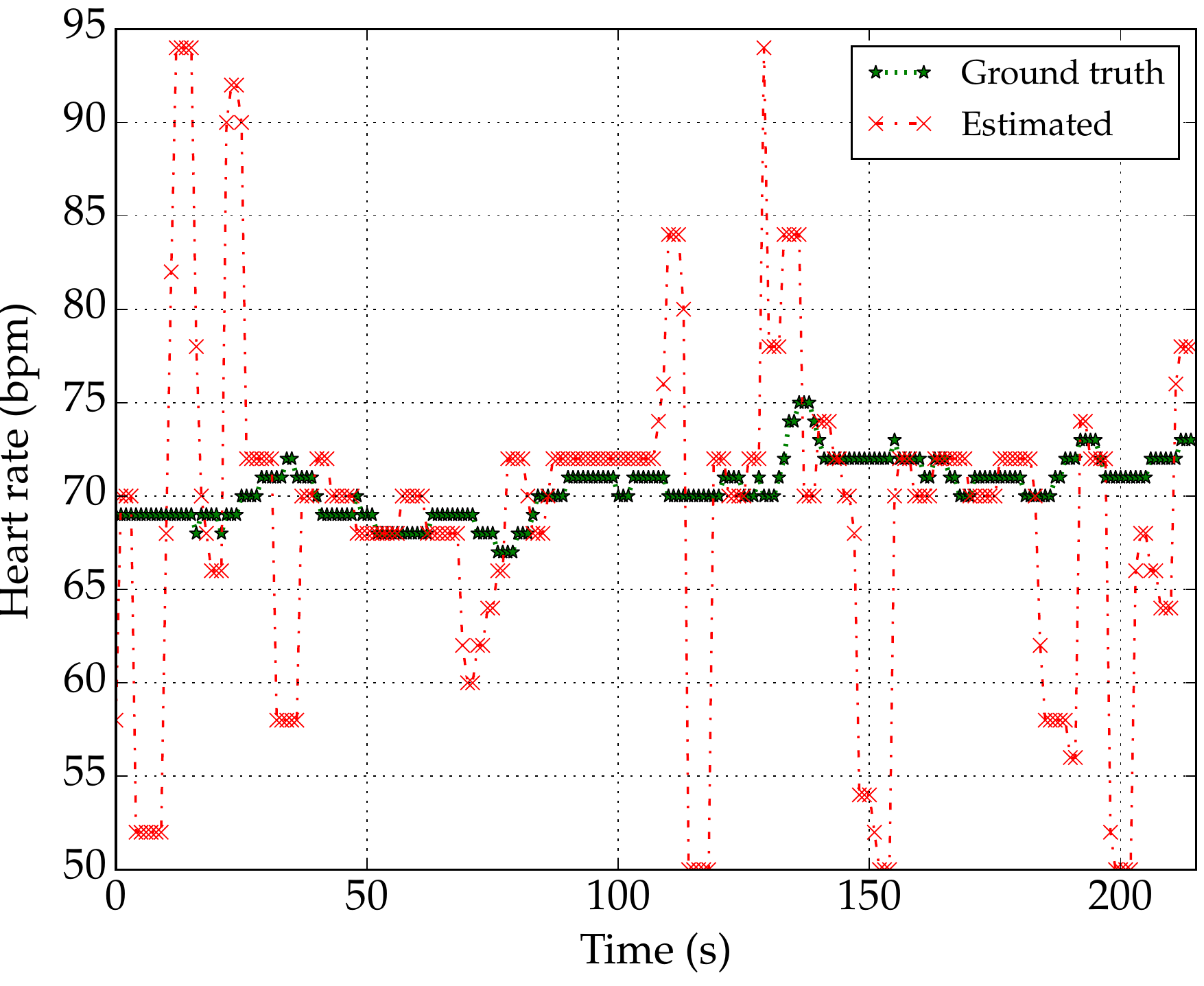}
  
  \caption{Low window size results in frequent, large jumps in heart rate estimates. The plot shows heart rate estimate for sitting user 2 using a window size of 10 seconds. }
  \label{fig:hr_low_wind}
  \end{center}
\end{figure}

\begin{figure}[tbhp]
\begin{center}
  \includegraphics[width=0.9\columnwidth]{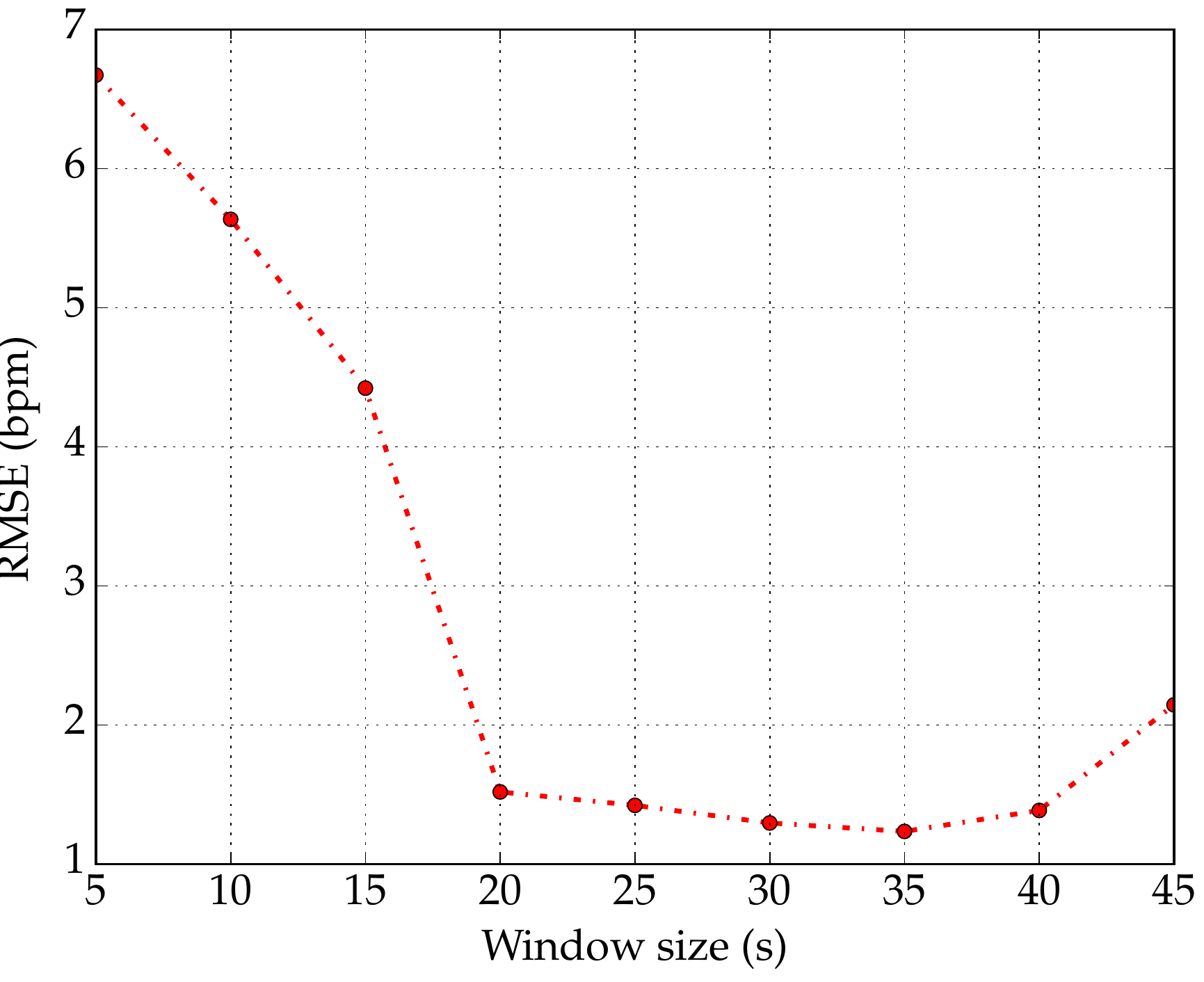}
  \caption{Effects of window size on heart rate estimation}
  \label{fig:hrwindow}
  \end{center}
\end{figure}

The procedures in estimating heart rate in real time from raw RSS measurements are summarized in Algorithm \ref{alg:heartrate}. The algorithm starts by initializing a buffer to store the incoming RSS samples continuously. At every second, mean-subtracted RSS data in the buffer is denoised using both Hampel and Butterworth bandpass filters. Then, the PSD of the resulting data is computed using the FFT. To improve pulse detection, we superimpose the PSD in resting cardiac frequency band (i.e. $f_{min}$ to $f_{max}$) upon the PSD in its second harmonic band (i.e., $2f_{min}$ to $2f_{max}$). The estimated heart rate corresponds to the frequency at which the maximum value of the sum of PSDs is obtained. To avoid inaccurate heart rate estimation as a result of random body motion, heart rate is estimated only when the peak value of the PSD is below a certain threshold as disturbance in the RSS data due to motion artifacts usually provide higher peak value of the PSD compared to the value obtained for pulse-induced signal.

\begin{algorithm}
\caption{Heart rate estimation algorithm} \label{alg:heartrate}
\begin{algorithmic}[1]
\BState{\textbf{Initialize:}}{}
\State $ count := 0$
\State $ RSSBuf := \text{buffer}(winSize)$

\BState{\textbf{Estimate HR:}}{}
\While{$true$}
    \State $mRSSI:= RSS - \text{mean}(RSS)$
    \State $RSSBuf:= \text{addToBuffer}(mRSSI)$
    \State $count = count+1$
     
    \If{$ count \geq f_s$}
            
        \State $count = 0$
        \State $filteredRSS:= \text{denoise}(RSSBuf)$
        \State $RSSPSD(f) := \text{PSD}(filteredRSS , f_s)$
        \State $PSD_1(f) := RSSPSD(f), \qquad f \in \{ f_{min}, f_{max}\}$
        \State $PSD_2(f) := RSSPSD(2f)\downarrow_2,\quad f\in \{ f_{min}, f_{max}\}$
        \State $sumPSD(f) := PSD_1(f) + PSD_2(f) $
        		   
        \If{$ \text{max}(sumPSD) < THD$} 
            \State $hRate := \underset{f}{\text{arg max}}\ sumPSD(f)$        
        \EndIf	
    \EndIf
\EndWhile

\end{algorithmic}
\end{algorithm}

\subsection{Experiment}
To evaluate our system for real time heart rate monitoring, we test three different subjects and two different environments. All subjects are healthy adults in the age between 24 to 29 years old. 

\begin{figure}[b]
    \begin{center}
  \begin{subfigure}[t]{0.25\textwidth}
     \includegraphics[width=\textwidth]{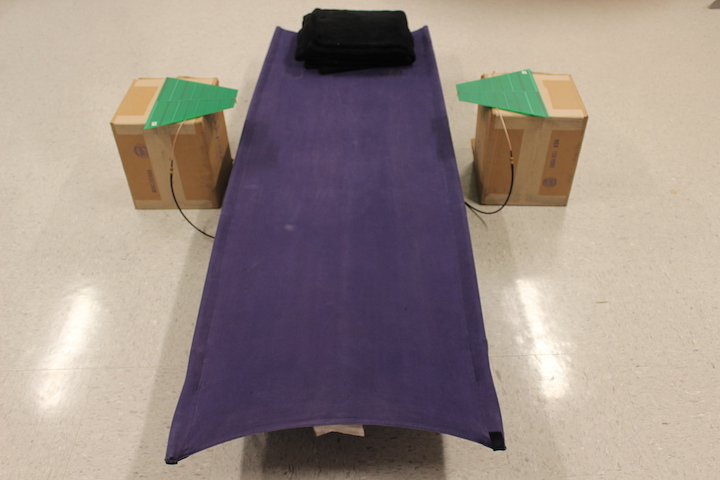}
  \caption{Lying on a cot}
  \label{fig:setup_lab}
  \end{subfigure}
  \hfill
  \begin{subfigure}[t]{0.215\textwidth}
     \includegraphics[width=\textwidth]{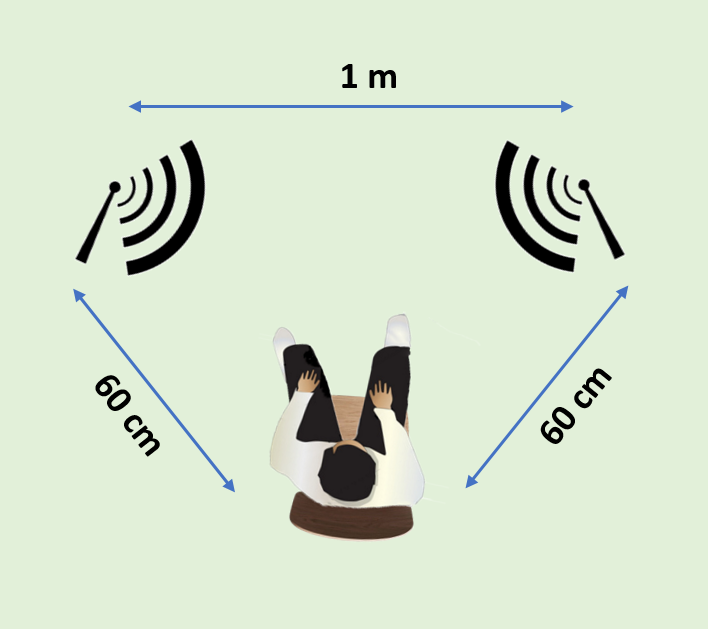}
   \caption{Sitting on a chair}
   \label{fig:setup_conf} 
    
  \end{subfigure}

  \caption{Experiment setup for heart rate monitoring.}
  \label{fig:setup}
  \end{center}
\end{figure}

The two environments are: 1) a research laboratory, and 2) a conference room.  Both are occupied with their typical furnishings, including chairs, desks, and equipment.  Both rooms have an approximate area of 56 m$^{\textnormal{2}}$, and only a single user is present in the room for each experiment. Fig.\ \ref{fig:setup_lab} shows the setup for experiments when the subject lies down. A cot elevates the person 15~cm above the floor, and the two directional antennas are set at 30~cm from the ground,  separated by 1~m from each other, and directed at the chest of the subject. In another setup, a subject sits in a chair 50~cm above the ground.  In this case, each antenna is 60~cm away from the chest of the subject, 1~m away from the other antenna, and 1~m above the ground as shown in Fig.\ \ref{fig:setup_conf}.  The receiver node of our system is connected to a laptop that processes the RSS data, and outputs heart rate estimates in real-time. 

To record the ground truth, we capture pulse rate measurements using a pulse oximeter physically attached to the subject's finger.  We use  Philips RespironicsNM3 respiratory profile monitor which is also connected to the same laptop. For this evaluation, we run each experiment for five minutes.  

\subsection{Results}

\begin{figure}
  \begin{center}
  \begin{subfigure}[t]{0.22\textwidth}
     \includegraphics[width=\textwidth]{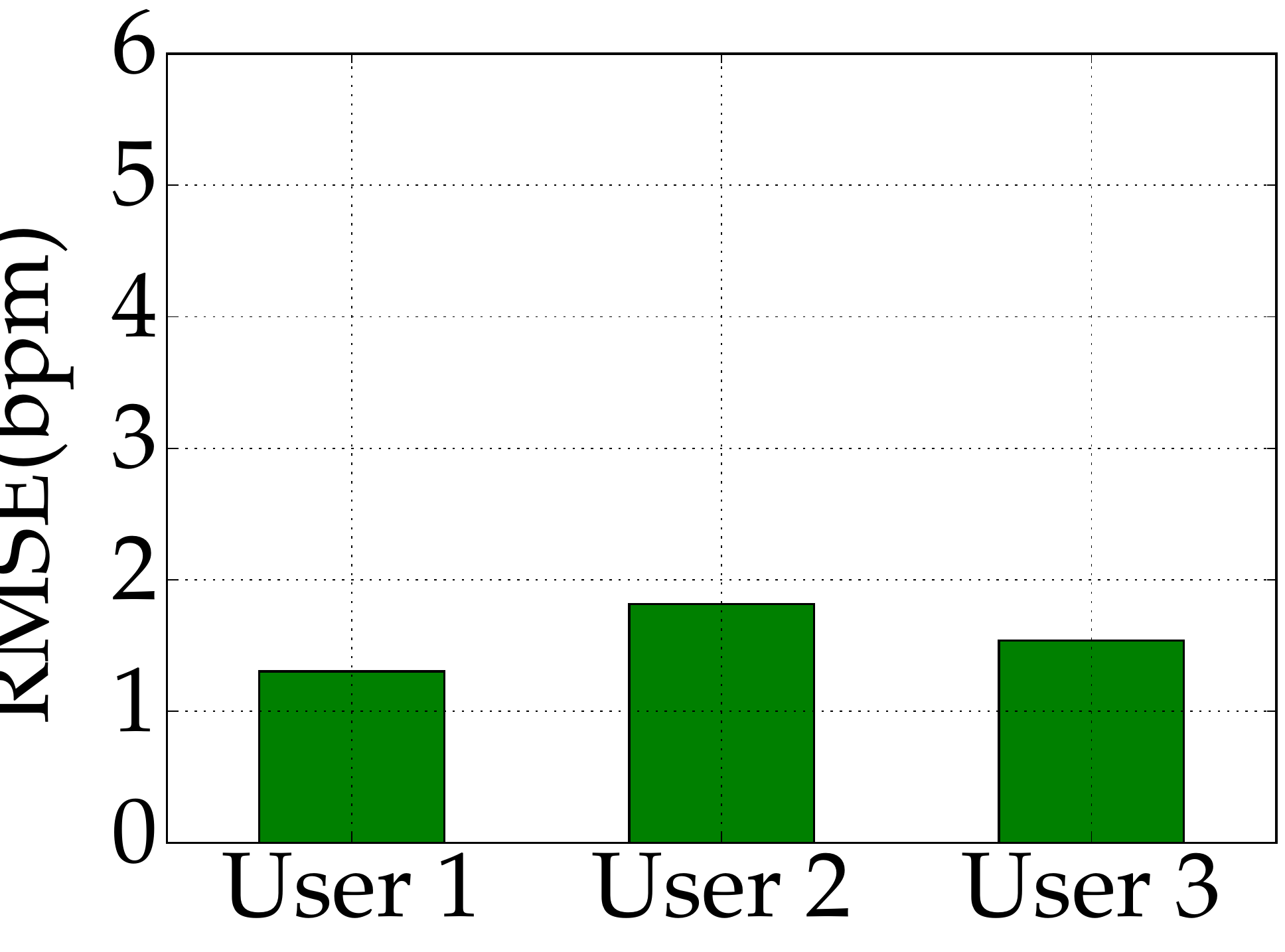}
  \caption{Heart rate RMSE: Lying on a cot}
  \label{fig:hr_lab}
  \end{subfigure}
  \hfill
  \begin{subfigure}[t]{0.22\textwidth}
     \includegraphics[width=\textwidth]{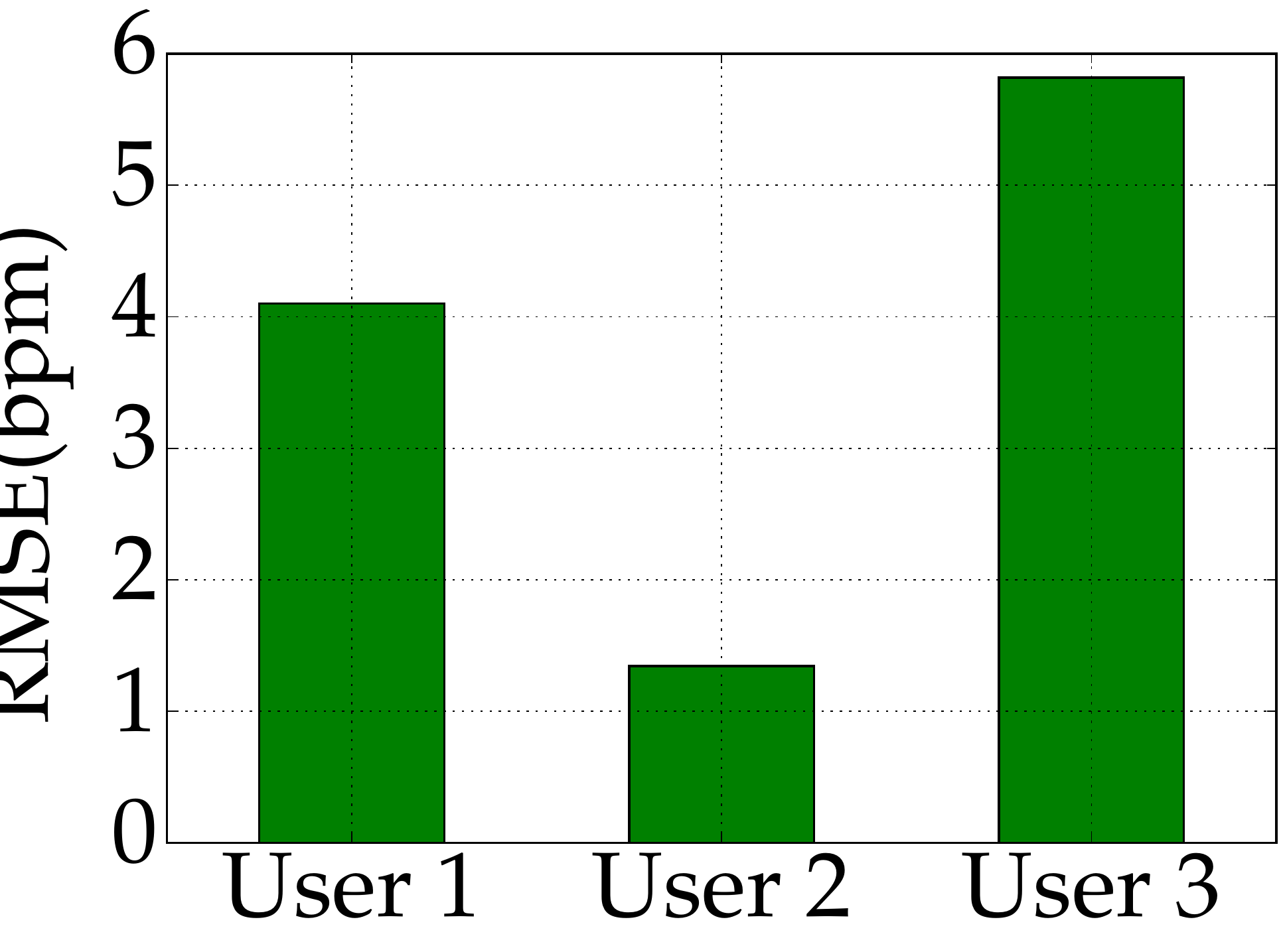}
   \caption{Heart rate RMSE: Sitting on a chair}
   \label{fig:hr_conf} 
    
  \end{subfigure}
  
  \label{fig:users}
  \caption{Heart rate RMSE for different subjects}
\end{center}
\end{figure}

We evaluate the performance of heart rate estimation in two different setups. The first setup involves a test in a conference room of size approximately $7~m\times8~m$ in which the subject sits on a chair, and the directional antennas separated by 1~m are set parallel to their chest. Fig.\ \ref{fig:hr_conf} shows the root mean squared error (RMSE) of heart rate estimates for each user. In Fig.\ \ref{fig:hr_lab}, we show the RMSE for lab setup where a subject lying on their back on a cot 15~cm above the ground. We observe that the RMSE for the second setup yields an RMSE of 1.57 bpm compared to 3.67 bpm for sitting users in the first setup.  It may be that the geometry for monitoring pulse via RF is better in the setup with the cot than with the chair.  However, it may also be that it is more difficult to be genuinely still while seated vs.\ when lying down.

We also note that the accuracy of real-time heart-rate estimation depends on the window size chosen to process the RSS signal. Fig.\ \ref{fig:hrwindow} shows the RMSE of heart-rate estimation averaged over three subjects as a function of processing window size for the lying down setup. The RMSE is shown to decrease with lower window size, but increases with higher widow size. Higher window size is not able to track the variability in heart rate while the performance of using lower window size is reduced as a result of lower SNR.

\begin{figure}[tbhp]
  \begin{center}
  \includegraphics[width=\columnwidth]{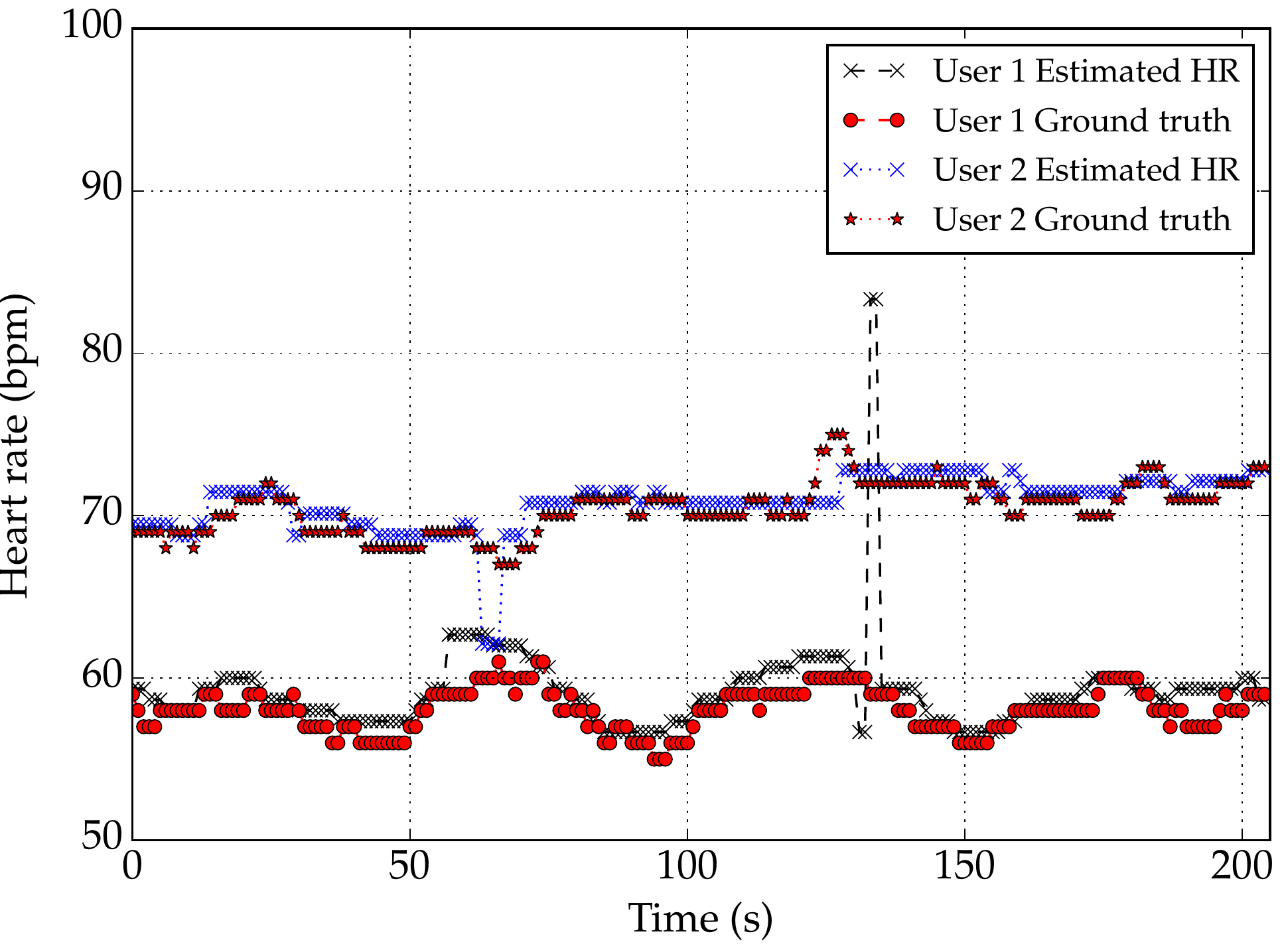}
  \caption{Estimated heart rate vs ground truth over time.  User 1 is lying on a cot, User 2 sitting on a chair.}
  \label{fig:hrtrack}
  \end{center}
\end{figure}

Fig.\ \ref{fig:hrtrack} quantifies the ability of the system to track the changes in heart rate over time.  Our algorithm is able to estimate the heart rate for users who have different heart rates, in the 67-73 bpm range for user sitting on a chair and in the 56-64 bpm range for user lying on a cot.  Also, the heart rate shown in Fig.\ \ref{fig:hrtrack} typically tracks the increases and decreases in heart rate observed over the course of tens of seconds for the person.  However, there are short periods during which the pulse rate estimate jumps very high or very low.  Further investigation will be required to track other motions of a person to determine if the jumps are caused by larger motion of their body, or if there is another cause.  Further work could also investigate tracking methods that could avoid the large jumps in the heart rate data. 

Given the 69 cm wavelength at the center frequency in use (434 MHz), and the fact that heart rate-induced skin vibrations may move the skin on the order of mm, it is quite surprising to be able to observe a person's heart rate in the measured RSS. However, the results indicate that RSS-based heart rate monitoring is possible with 20 s of data, that it can track heart rate variations over time, and that it can stay within 1.6 bpm RMSE of the rate measured by a clinical-grade pulse oximeter.

\section{Gesture Recognition}
\label{sec:gesture}

RF based gesture recognition exploits the change in wireless channel information as a result of human limb and hand movements \cite{pu2013whole, sigg2014telepathic, wang2015understanding}. In this section, we present the results of using our system for gesture recognition. Our results demonstrate that accurate gesture recognition is possible using such a narrow bandwidth and low transmit power.  

Prior work in gesture recognition has demonstrated recognition of gestures using RF measurements, but has required use of a WiFi channel bandwidth \cite{pu2013whole,sigg2014telepathic}, or use of software-defined radio hardware \cite{pu2013whole,kellogg2014bringing}.  Our work shows that the same number of gestures can be recognized as other related work, with a relatively high accuracy.  We note that \cite{kellogg2014bringing} implements a compelling low-power narrowband analog platform for gesture recognition, but that the results of the paper on eight-gesture recognition are performed using a USRP-based prototype.  In comparison, our system is even lower in bandwidth utilization and uses less transmit power, and is fully implemented using a BBG and TI CC1200.  We note that except for \cite{sigg2014telepathic}, each system fully occupies the channel by transmitting 100\% of the time.    We compare system specifications and reported experimental classification accuracies in Table \ref{table:gestures}.  

Although \cite{wang2015understanding} is not included in Table \ref{table:gestures} because it is an activity recognition paper, we note it uses CSI measurements over a 20 MHz bandwidth on a 3$\times$3 MIMO channel to classify nine different activities.  When it uses a WiFi packet rate of 400 Hz, it achieves an accuracy of 87\%.

\begin{table*}[tbhp]
\centering
\begin{tabular}{ |l|c|c|c|c|c|c|c| } 
 \hline
 \bf Paper & \bf BW (MHz) & \bf \# Antennas & \bf Hardware & \bf TX Power &\bf \# Gestures & \bf Dist & \bf Accuracy \\ [0.5ex] 
 \hline
 Pu et al. \cite{pu2013whole}  
       & 10       & 5 & USRP-N210  & 10 mW  & 9 & $<$ 5 m & 94\% \\ 
 Sigg et al. \cite{sigg2014telepathic} 
       & 20       & ? & Phone WiFi & ?      & 7 & $<$ 1 m & 56\% \\ 
 Kellogg et al. \cite{kellogg2014bringing}            
       & 0.1-0.2  & 1 & USRP       & 346 mW & 8 & $<$ 1 m & 97\% \\  
 This paper 
       & 0.01    & 1 & TI CC1200  & 1 mW   & 8 & 1.5 m   & 85\% \\
 \hline
\end{tabular}
\caption{Related work on gesture recognition: Hardware specs, reported experimental accuracies}
\label{table:gestures}
\end{table*}

RSS-based SISO approaches like \cite{sigg2014telepathic, kellogg2014bringing} yield relatively low cost, efficient bandwidth utilization, and low computational complexity.  However, RSS for a narrowband channel is a single dimensional measurement; it is unable to simultaneously provide the measurements from multiple OFDM subcarriers and multiple pairs of MIMO antennas as in \cite{wang2015understanding,pu2013whole}. We address this challenge by modifying the approach of \cite{wang2015understanding} to work with single-channel RSS data.  In this section, we present a gesture recognition system that makes use of high resolution RSS estimates on a single narrowband channel to detect eight different gestures, results which we believe are useful  for studying the design space for future gesture recognition systems.

\subsection{Methods}
We apply DWT based classification for gesture recognition, which is performed in two phases: training and testing. In the training phase, labelled RSS data is used to train a machine learning classifier. In the testing cycle, a test data is classified into one of the trained gestures. Each cycle involves three steps. First, the gesture regions in the RSS data are identified using variance based segmentation after the data is preprocessed. Second, for each RSS segment identified as a gesture, we perform time-frequency analysis to extract features. Finally, we train classifiers with labelled data or classify incoming test data.

\subsubsection{Prepossessing and Segmentation}
The noisy RSS data is preprocessed and filtered to reduce the effect of outliers and additive noise. We apply a Hampel filter \cite{pearson2016generalized} to replace outliers in the data, and a Butterworth low-pass filter to reduce additive noise. Since some gestures induce high-frequency RSS signal, we set the cutoff frequency for the low-pass filter to 90~Hz.  

Next, segmentation is implemented to detect motion and identify time windows in which a gesture is performed. Since motion increases the variance of such RSS data, we apply moving variance-based thresholding on locally mean-removed RSS data to detect motion and thus extract RSS segments containing a gesture, as follows. At a given instant of time $t$, the variance of a short-term buffer of RSS is calculated.  These variance values are buffered and the maximum over a long-term period is computed.  A gesture is detected whenever the ratio of the current RSS variance to the long-term maximum variance exceeds a threshold $T$ for at least 0.2 seconds.

\subsubsection{Feature Extraction}
In this step, properties that differentiate the incoming RSS segment data are selected and computed. 

\begin{figure*}[!htb]
  \begin{center}
  \begin{subfigure}[t]{0.32\textwidth}
    \includegraphics[width=\linewidth]{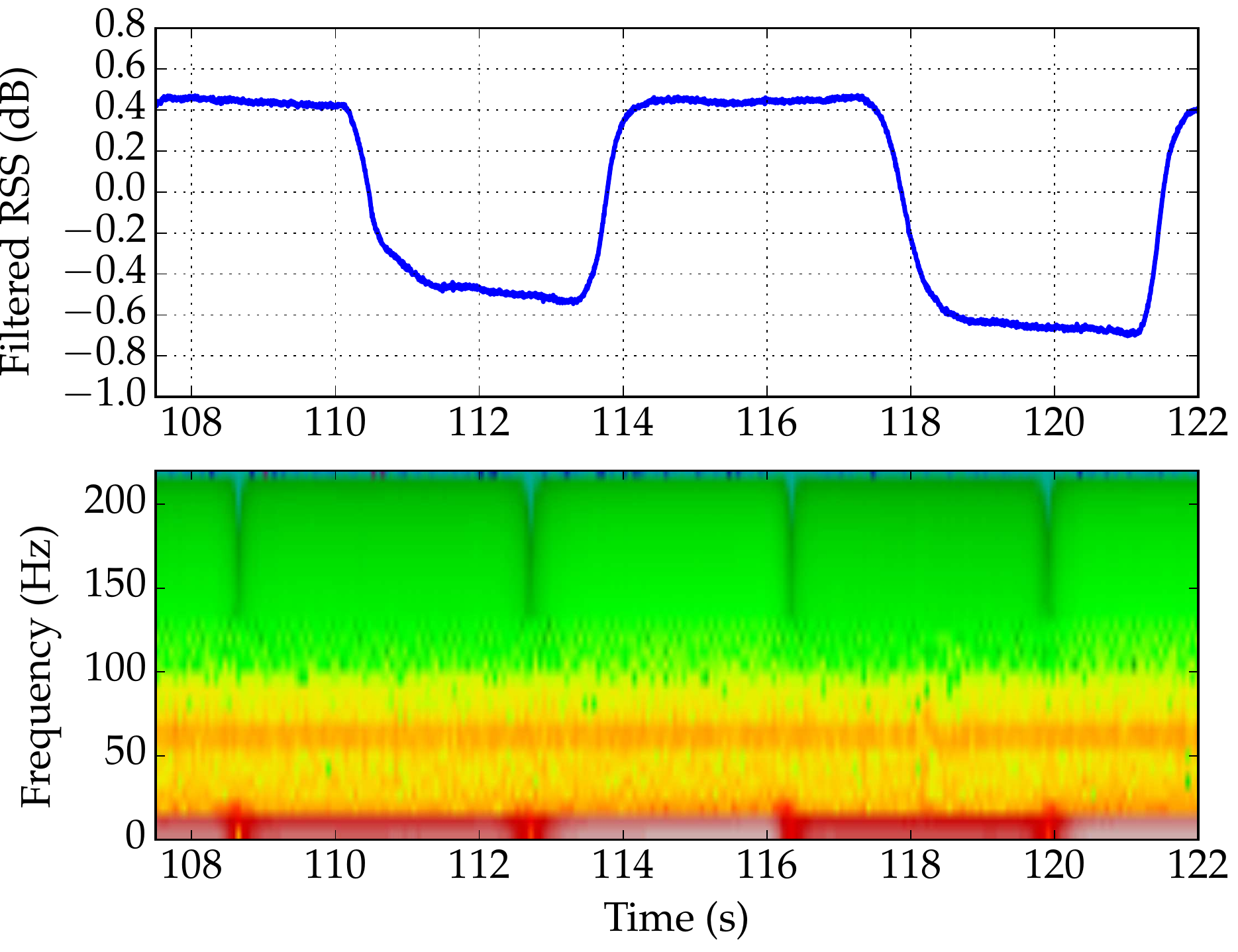}
    \caption{Push \& Pull}
  \end{subfigure}
  \hfill
  \begin{subfigure}[t]{0.32\textwidth}
    \includegraphics[width=\linewidth]{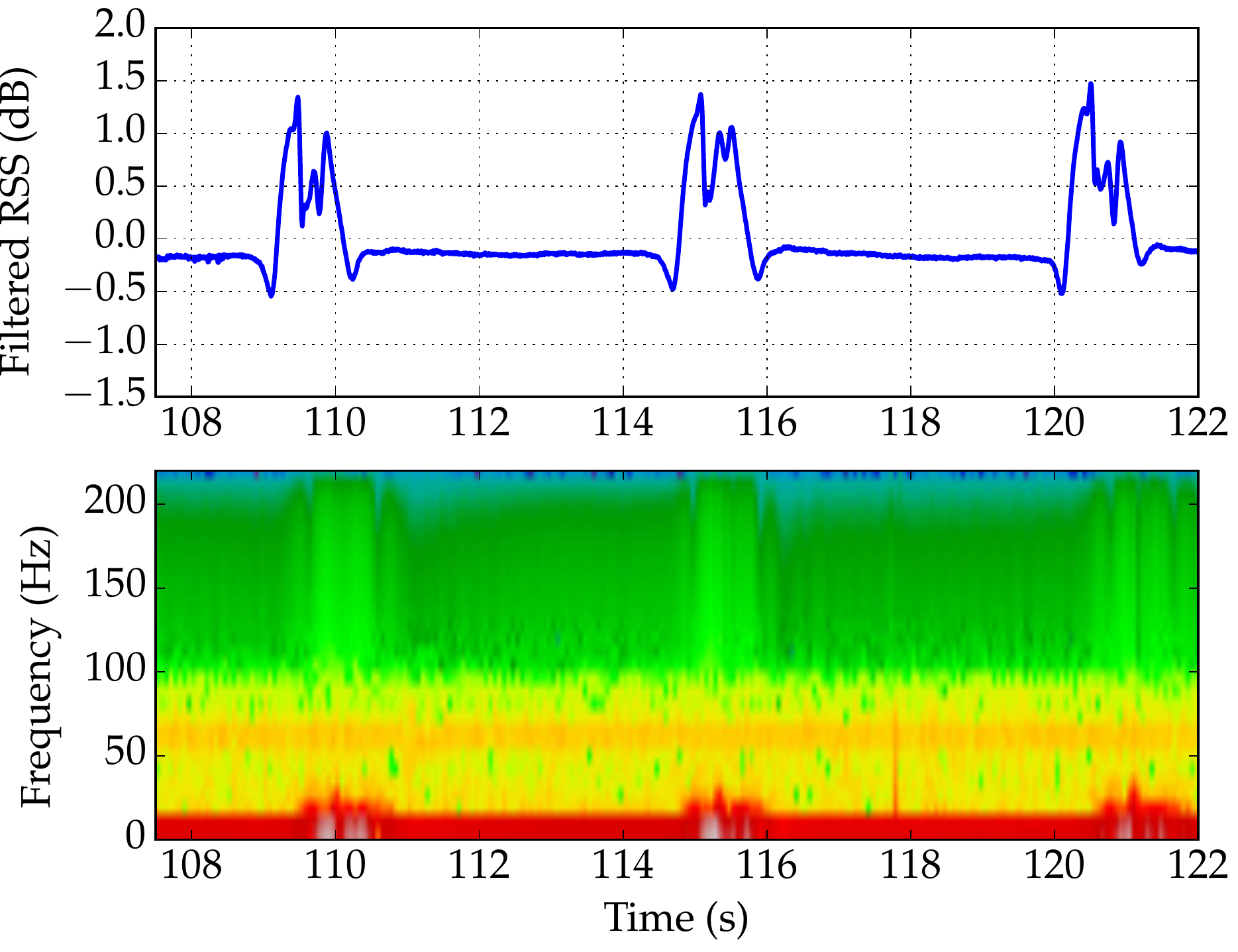}
    \caption{Kick}
    
  \end{subfigure}
  \hfill
   \begin{subfigure}[t]{0.32\textwidth}
    \includegraphics[width=\linewidth]{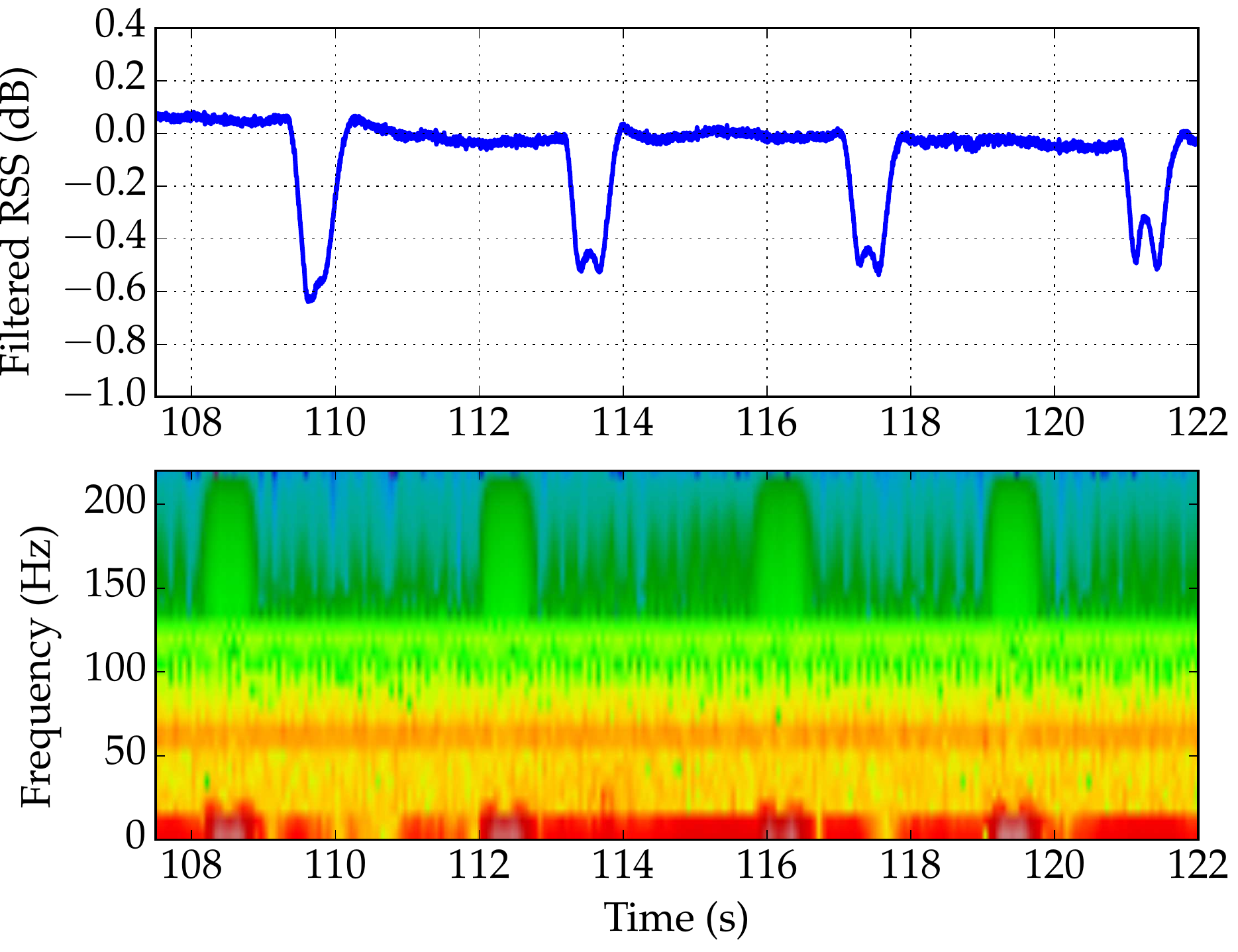}
    \caption{Punch}
    
  \end{subfigure}
  \caption{RSS waveforms and spectrogram for three gestures: Each has unique features in time and frequency domains. }  \label{fig:gestures}
\end{center}
\end{figure*}

First, each RSS segment detected as a gesture is  processed to compute the DWT, which provides multiple resolutions of time and frequency-domain characteristics of the RSS signal \cite{wang2015understanding}. As an example, the spectrogram of the RSS data for three different gestures is shown in Fig.\  \ref{fig:gestures}, which shows that the RSS data provides unique features both in time and frequency domains. We use the \emph{Daubechies 3} wavelet for analysis, which showed high accuracy across our numerous tests.

Next, selected features are extracted from each RSS segment. We select features that provide high variation across different types of gestures, and small variance within each gesture type. Selected features include mean, variance, maximum, minimum, maximum energy, average frequency, and half point frequency. The features are computed at the first three DWT decomposition levels.  We decompose each RSS segment into components at four different DWT bands. The DWT components considered in this study are the detail coefficients at the first three decomposition levels (i.e., cD1, cD2, cD3), and approximation coefficients at level three (i.e., cA3).  Selected features are computed for each DWT component extracted. Average frequency and half point frequency are determined from the power spectral density of the given DWT component. In addition to features computed at different DWT bands, the feature set also includes DWT approximation (cA3) and detail (cD3) coefficients at level 3. With this process, a total of $671$ features are extracted from each RSS segment.  Compared to \cite{wang2015understanding}, we have selected a very different set of DWT-based features, as the important characteristics of our gestures are different from the activities as studied in \cite{wang2015understanding}, for which speed is critical.

\subsubsection{Classification}

Finally, the extracted features are used to classify the gesture type.  
We consider and compare the performance of three different machine learning classifiers: $k$-nearest neighbors (KNN), support vector machines (SVM) and random forest classifiers. Specifically, we use KNN with five neighbors, SVM with a linear kernel and random forest with $15$ estimators. The parameters of the classifiers are selected for their improved accuracy across multiple tests without significantly affecting the computational requirements. In the training phase, each classifier is trained with features extracted from labelled training data and gesture labels. During the test phase, an unlabelled gesture's data is input to the classifier and an estimated label is output.

\subsection{Experiment}
The experiments for gesture recognition are conducted in two different settings, and involve three different subjects. We consider eight different gestures as defined in \cite{pu2013whole}, namely \emph{punch}, \emph{punchx2}, \emph{kick}, \emph{strike}, \emph{drag}, \emph{dodge}, \emph{push}, and \emph{pull}. 
We conduct the experiments in empty office and research laboratory settings. The laboratory has an approximate area of 56~m$^{\textnormal{2}}$, and is furnished with desks, chairs, and other equipment while the empty office has an approximate area of 17~m$^{\textnormal{2}}$. In both settings, the nodes are set at a height of 75~cm above the ground, and  2.28~m apart from each other. For this experiment, we use omnidirectional antennas.  
Each user performs a gesture at a perpendicular distance of 1.5~m from the link line while facing  the link line. We record 512 training gestures and 328 testing gestures from the three different users. 

\subsection{Results}

\begin{figure}[tbhp]
\begin{center}
  \includegraphics[width=\columnwidth]{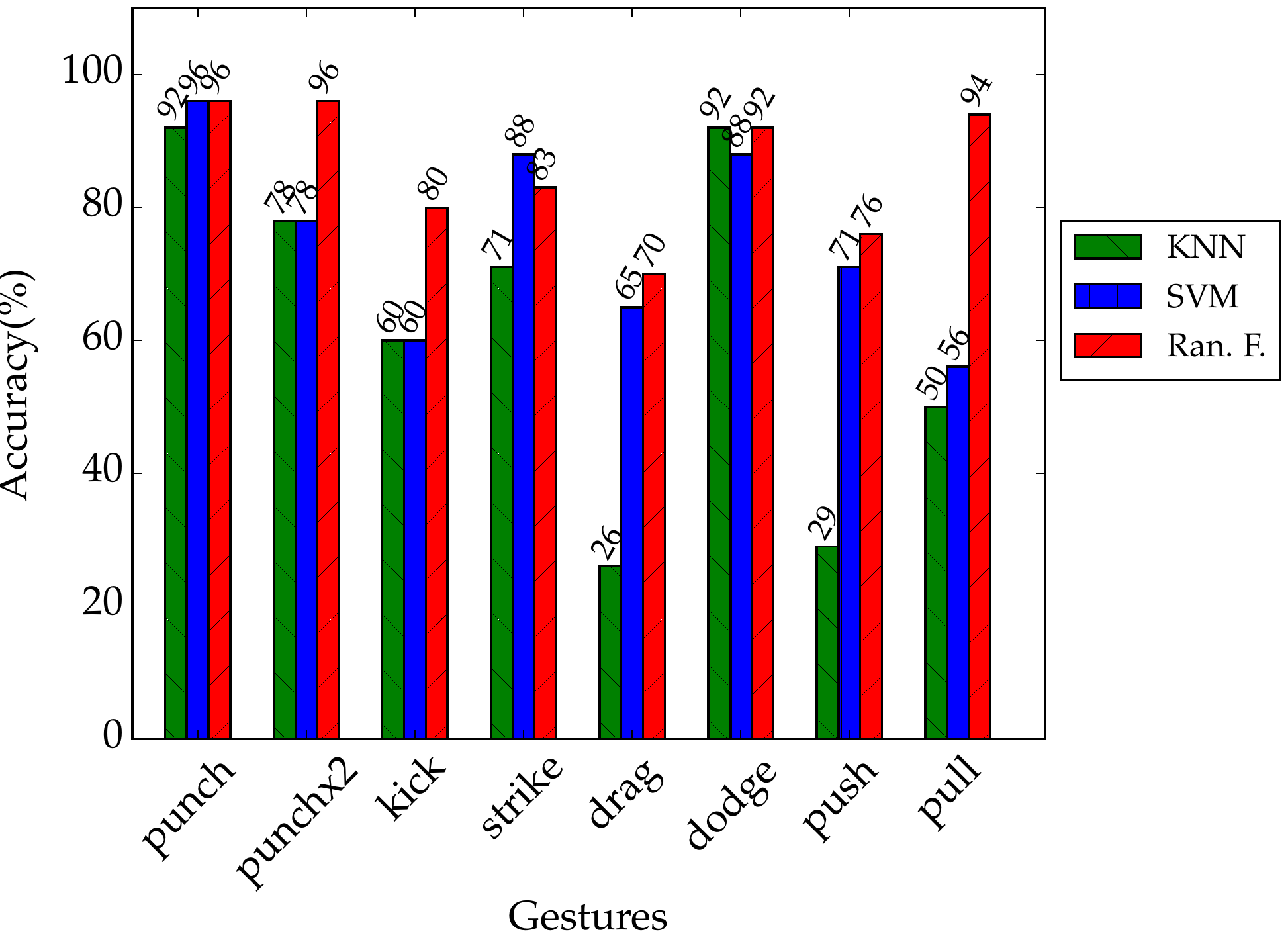}
  \caption{Comparison of classifiers: with random forest classifier,  more than 90\% correct prediction is obtained for half of the gestures}
  \label{fig:classifiers}
  \end{center}
\end{figure}

We evaluate the performance of our system in classifying gestures. We collect and label RSS data for each measurement. We apply and compare three classifiers, namely KNN, SVM and random forest classifiers. Fig.\  \ref{fig:classifiers} shows the accuracy obtained with these classifiers. K-nearest neighbors performs poorly at average accuracy of 63\% while random forest outperforms the other classifiers attaining an average accuracy of 85\%. This may indicate that decision-tree based classification provides improved gesture recognition accuracy for selected set of features as compared to instance based classification. We show the confusion matrix for random forest classifier in Table \ref{tab:confusion}. We observe that some gestures such as \emph{drag} result in a lower accuracy due to similarity in their RSS waveform with other  gestures. However, for half of the gestures, our system provides accuracy above 90\%.  

In  summary, we show the capability of high resolution RSS measurement in gesture recognition. Compared to \cite{sigg2014telepathic}, which attained an accuracy of 56\% for seven gestures, and compared to \cite{luong2016rss} which attained an accuracy of  83\% for four gestures, we demonstrate that applying DWT based classification to  fine-grained RSS data provides a higher accuracy of 85\% even for eight different gestures.

\begin{table*}[]
\centering
\caption{Confusion matrix for random forest classier}
\label{tab:confusion}
\begin{tabular}{clcccccccc}
\cline{3-10}
\multirow{8}{*}
{\rotatebox[origin=c]{90}{\textbf{Predicted}}} & \multicolumn{1}{r|}{punch} & \multicolumn{1}{c|}{\cellcolor{green} 0.96} & \multicolumn{1}{c|}{0.04} & \multicolumn{1}{c|}{0.00} & \multicolumn{1}{c|}{0.00} & \multicolumn{1}{c|}{0.09} & \multicolumn{1}{c|}{0.00} & \multicolumn{1}{c|}{0.06} & \multicolumn{1}{c|}{0.00} \\ \cline{3-10} 
 & \multicolumn{1}{r|}{punchx2} & \multicolumn{1}{c|}{0.04} & \multicolumn{1}{c|}{\cellcolor{green} 0.96} & \multicolumn{1}{c|}{0.00} & \multicolumn{1}{c|}{0.00} & \multicolumn{1}{c|}{0.00} & \multicolumn{1}{c|}{0.00} & \multicolumn{1}{c|}{0.00} & \multicolumn{1}{c|}{0.00} \\ \cline{3-10} 
 & \multicolumn{1}{r|}{kick} & \multicolumn{1}{c|}{0.00} & \multicolumn{1}{c|}{0.00} & \multicolumn{1}{c|}{\cellcolor{green} 0.80} & \multicolumn{1}{c|}{0.00} & \multicolumn{1}{c|}{0.04} & \multicolumn{1}{c|}{0.04} & \multicolumn{1}{c|}{0.00} & \multicolumn{1}{c|}{0.00} \\ \cline{3-10} 
 & \multicolumn{1}{r|}{strike} & \multicolumn{1}{c|}{0.00} & \multicolumn{1}{c|}{0.00} & \multicolumn{1}{c|}{0.00} & \multicolumn{1}{c|}{\cellcolor{green} 0.83} & \multicolumn{1}{c|}{0.00} & \multicolumn{1}{c|}{0.00} & \multicolumn{1}{c|}{0.18} & \multicolumn{1}{c|}{0.00} \\ \cline{3-10} 
 & \multicolumn{1}{r|}{drag} & \multicolumn{1}{c|}{0.00} & \multicolumn{1}{c|}{0.00} & \multicolumn{1}{c|}{0.08} & \multicolumn{1}{c|}{0.00} & \multicolumn{1}{c|}{\cellcolor{green} 0.70} & \multicolumn{1}{c|}{0.00} & \multicolumn{1}{c|}{0.00} & \multicolumn{1}{c|}{0.00} \\ \cline{3-10} 
 & \multicolumn{1}{r|}{dodge} & \multicolumn{1}{c|}{0.00} & \multicolumn{1}{c|}{0.00} & \multicolumn{1}{c|}{0.12} & \multicolumn{1}{c|}{0.00} & \multicolumn{1}{c|}{0.17} & \multicolumn{1}{c|}{\cellcolor{green} 0.92} & \multicolumn{1}{c|}{0.00} & \multicolumn{1}{c|}{0.06} \\ \cline{3-10} 
 & \multicolumn{1}{r|}{push} & \multicolumn{1}{c|}{0.00} & \multicolumn{1}{c|}{0.00} & \multicolumn{1}{c|}{0.00} & \multicolumn{1}{c|}{0.17} & \multicolumn{1}{c|}{0.00} & \multicolumn{1}{c|}{0.00} & \multicolumn{1}{c|}{\cellcolor{green} 0.76} & \multicolumn{1}{c|}{0.00} \\ \cline{3-10} 
 & \multicolumn{1}{r|}{pull} & \multicolumn{1}{c|}{0.00} & \multicolumn{1}{c|}{0.00} & \multicolumn{1}{c|}{0.00} & \multicolumn{1}{c|}{0.00} & \multicolumn{1}{c|}{0.00} & \multicolumn{1}{c|}{0.04} & \multicolumn{1}{c|}{0.00} & \multicolumn{1}{c|}{\cellcolor{green} 0.94} \\ \cline{3-10} 
\multicolumn{1}{l}{} & \multicolumn{1}{l}{} & punch & punchx2 & kick & strike & drag & dodge & push & pull \\
\multicolumn{1}{l}{} & \multicolumn{1}{l}{} & \multicolumn{8}{c}{\textbf{Actual}}
\end{tabular}
\end{table*}

\section{Human Speed Estimation}
\label{sec:speed}

In this section, we present and evaluate our system for the estimation of human walking speed from single-channel RSS measurements.  Our system is ``device-free,'' that is, it does not require the person to carry a wireless device.  One way to perform such human speed estimation is to first perform RF-based device-free localization \cite{yang2013rssi} and then derive the speed from the coordinate estimates over time; however, such systems require many transceivers deployed around the area of interest. A few other techniques directly measure speed or range using a radar measurement system \cite{van2008feature,adib20143d}; however, such systems use GHz of bandwidth.  In this paper, we propose speed estimation using a single pair of narrowband transceivers which operate on the spectral components of the measured RSS data.   

When a person crosses a wireless link, which we refer as a link line, disturbances in the wireless channel result in changes in the RSS measurements, as shown in Fig.\ \ref{fig:avgfreq}.  These changes provide information that enables an inexpensive way of estimating the speed of the person when they cross the link. 

One application would be for security purposes, to know how fast a person is travelling though an area. Another application may be in home health monitoring to be aware of signs of physical decline (that an elder may not be aware of themselves). Past research has used a depth camera to monitor changes in speed of the resident of a home \cite{stone2013unobtrusive} with the goal of predicting future falls.  Our system could have similar purposes without use of a vision-based sensor, which people may object to for privacy reasons.

\subsection{Motivation for Approach}
\begin{figure}[tbhp]
\begin{center}
  \includegraphics[width=\columnwidth]{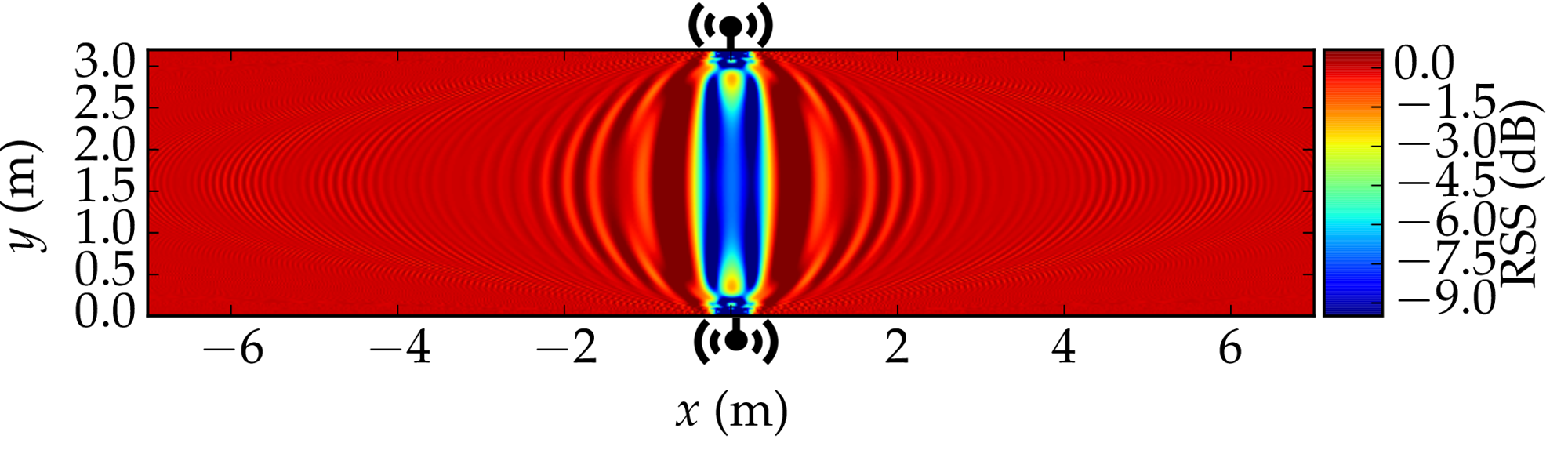}
  \caption{Simulated RSS as a function of location of a person based on diffraction based model in \cite{rampa2015physical}}
  \label{fig:rss_map}
  \end{center}
\end{figure}

When a person moves near a wireless link, the RSS changes due to shadowing induced by their presence.  Human shadowing nearby a wireless link can be explained using diffraction theory. Based the diffraction based model proposed by Rampa \emph{et al.} \cite{rampa2015physical}, we simulate the RSS as a function of a person's location with respect to the transceivers, and the results are shown in Fig.\ \ref{fig:rss_map}. Kaltiokallio \emph{et al.} also derive a three-state temporal RSS model in which RSS is related to position and velocity of a cylinder \cite{kaltiokallio2014enhancing}. From these models, it can be shown that the peak frequency of the RSS signal decreases as the cylinder approaches the link line.

\begin{figure}[tbp]
  \begin{center}
  \begin{subfigure}[t]{0.45\textwidth}
   \centering
    \includegraphics[width=0.65\textwidth]{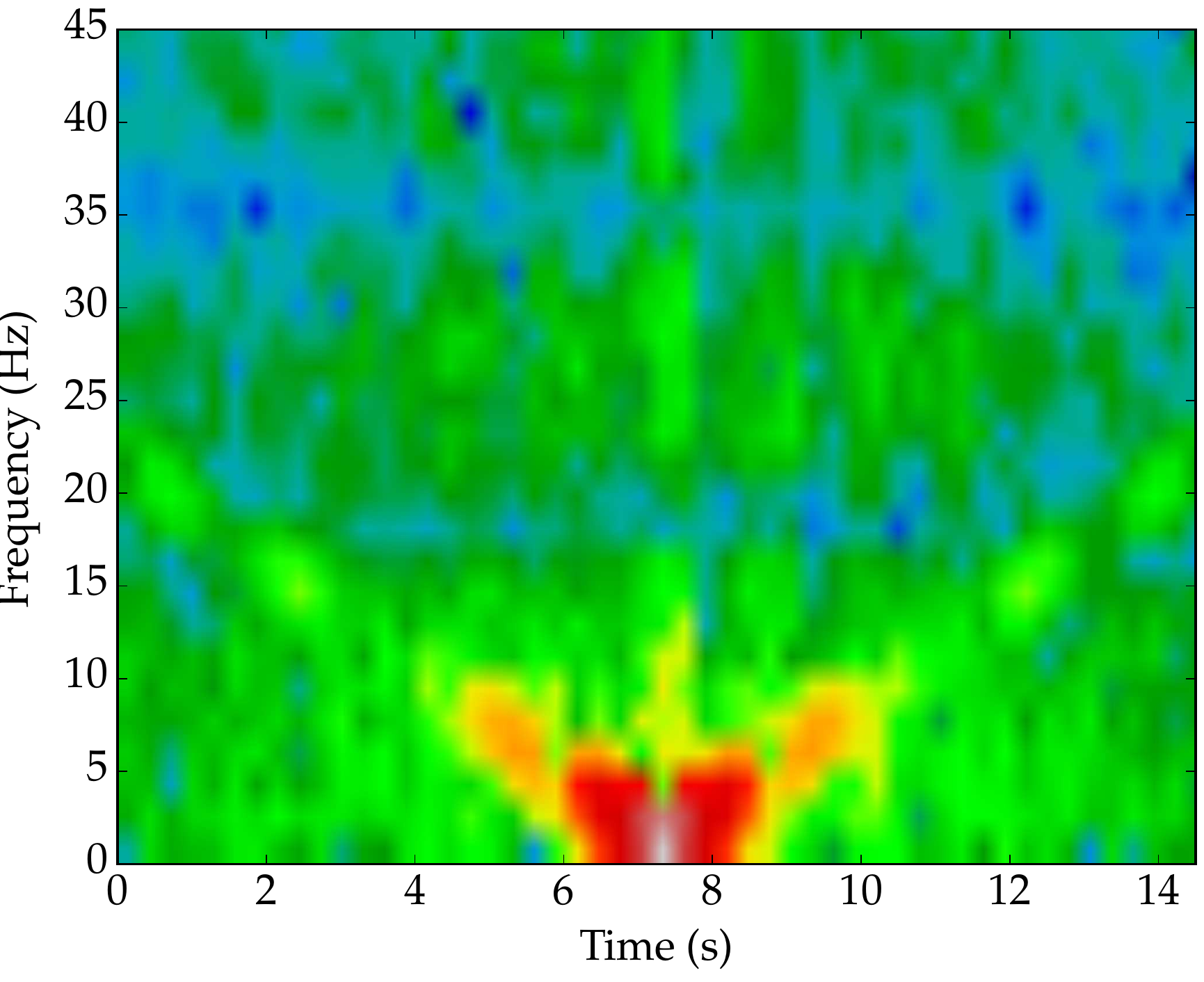}
    \caption{Simulated for cylinder crossing at $t=7$ seconds}
    \label{fig:sim_spec_cross}
  \end{subfigure}
  \hfill
  \begin{subfigure}[t]{0.45\textwidth}
    \centering
    \includegraphics[width=0.65\textwidth]{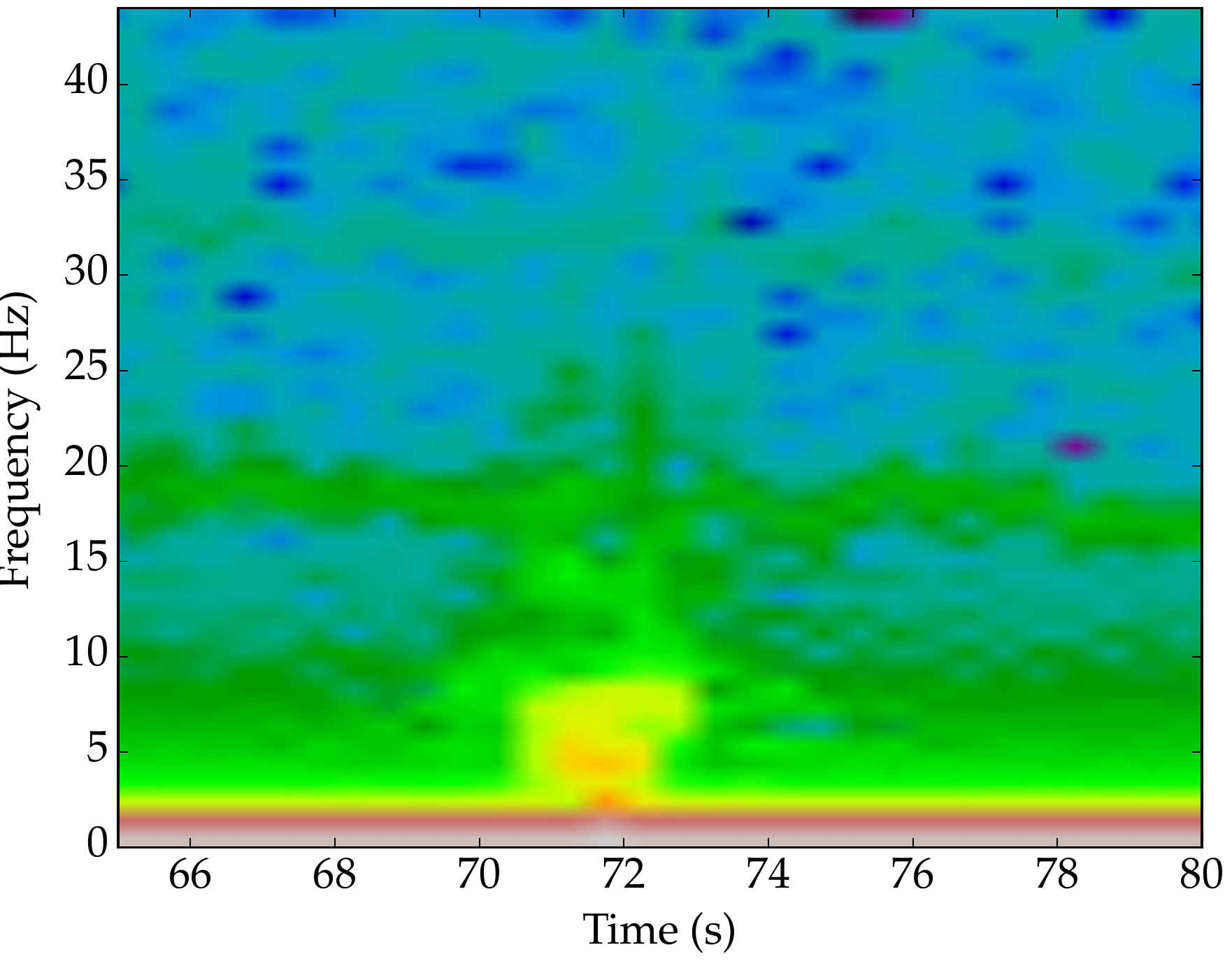}
    \caption{Measured from human crossing at $t=72$ seconds}
    \label{fig:real_spec_cross}
  \end{subfigure}
  
  \label{fig:spec_cross}
  \caption{Spectrogram of of (a) simulated and (b)measured RSS data. }
\end{center}
\end{figure}

We simulate line-crossing using the human-as-a-cylinder model of \cite{rampa2015physical}, and show the spectrogram of the resulting RSS in Fig.\ \ref{fig:sim_spec_cross}. It can be observed that the peak frequency decreases as the person approaches the link line. The lowest peak frequency occurs at the time of crossing.  This lowest peak frequency is directly proportional to the speed at the time of crossing.

However, actual human motion is unlike a cylinder as assumed in the model of \cite{rampa2015physical}. Different body parts such as the legs, torso and hands move at different instantaneous velocities, causing different Doppler shifts in the RF waves with which they interact.  In Fig.\ \ref{fig:real_spec_cross}, we show the spectrogram of the measured RSS during a time when an actual walking person crosses the link line. 

\subsection{Proposed Method}

\begin{figure}[tbhp]
\begin{center}
  \includegraphics[width=\columnwidth]{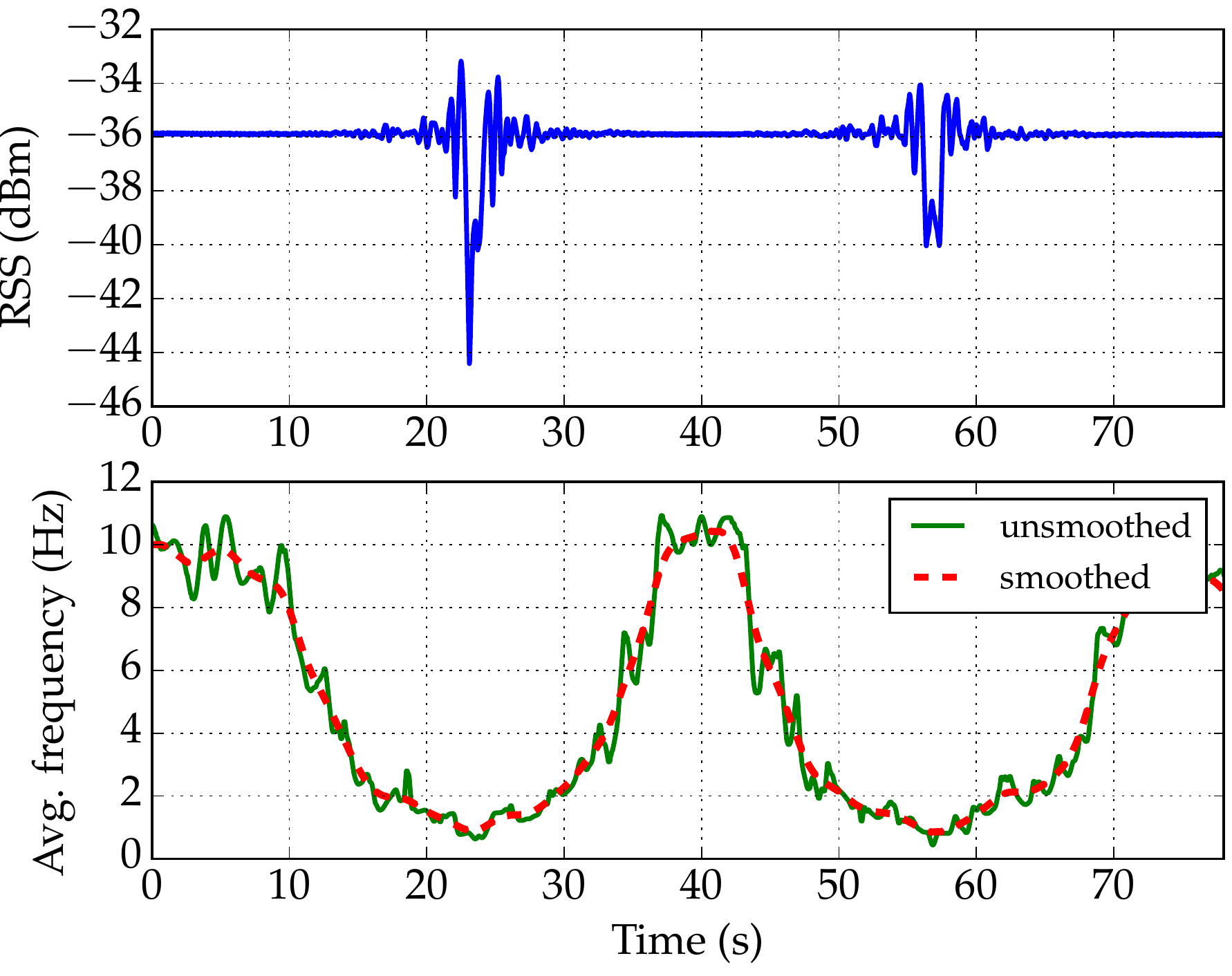}
  \caption{Average RSS frequency for measured line crossing with 2 second windows: The average frequency is smoothed using moving average filter before speed estimation. }
  \label{fig:avgfreq}
  \end{center}
\end{figure}

We propose instead that the average frequency of the RSS waveform can be used to detect human motion, and to estimate their average speed. The average frequency $f_{av}$ at time $t$ is calculated from the PSD of the locally mean-removed RSS as
\begin{equation}\label{eqn:avgfreq}
f_{av}(t) = \dfrac{\sum_{i} f_i S_r(f_i,t)}{\sum_{i} S_r(f_i,t) },
\end{equation}
where $S_r(f_i,t)$ is the power of the spectrogram of the RSS signal at frequency $f_i$ at time $t$. 
In our study, we consider human walking speeds in the range from 0.3~m/s to 1.8~m/s. For the speed range and a sampling rate of 449~Hz, average RSS frequency computed from a two-second sliding window provides accurate speed estimates. 
The estimated average frequency is smoothed using a moving-average filter before determining the average speed. Fig.\ \ref{fig:avgfreq} shows before and after smoothing the average frequency. It is noted that the minimum average frequency on the smoothed graph corresponds to the instant when the person crosses the link line.

{\bf Crossing Prediction.}
The frequency response of the RSS waveform provides information about a person who is approaching, even before they cross, a link line.  When a person is far away from a link line, the person has little effect on the RSS values.  The measured average frequency at this point is relatively high due wideband RSS noise, as shown in Fig.\ \ref{fig:avgfreq}. As the person approaches the link line the amplitude of oscillation increases, which translates to higher and higher power at a relatively low frequency in the spectrogram. At crossing, there is a a sudden drop in the RSS in addition to a noticeable decrease in the frequency of the oscillating signal. 
As a result, crossing prediction can be performed by thresholding the average frequency, i.e., when a person approaches a link line the average frequency drops below a certain threshold. Our algorithm starts operating the speed estimation algorithm when this threshold is crossed.  

{\bf Speed Estimation.}
Next, we estimate the speed of the person by scaling the $f_{av}$ at the time of crossing by a constant $\alpha$,
\[
\hat{v} = \alpha \min_{t} f_{av}(t),
\]
where the minimum is taken over a constant time interval.

\subsection{Experiment}
\begin{figure}[htbp]
\begin{center}
  \includegraphics[width=\columnwidth]{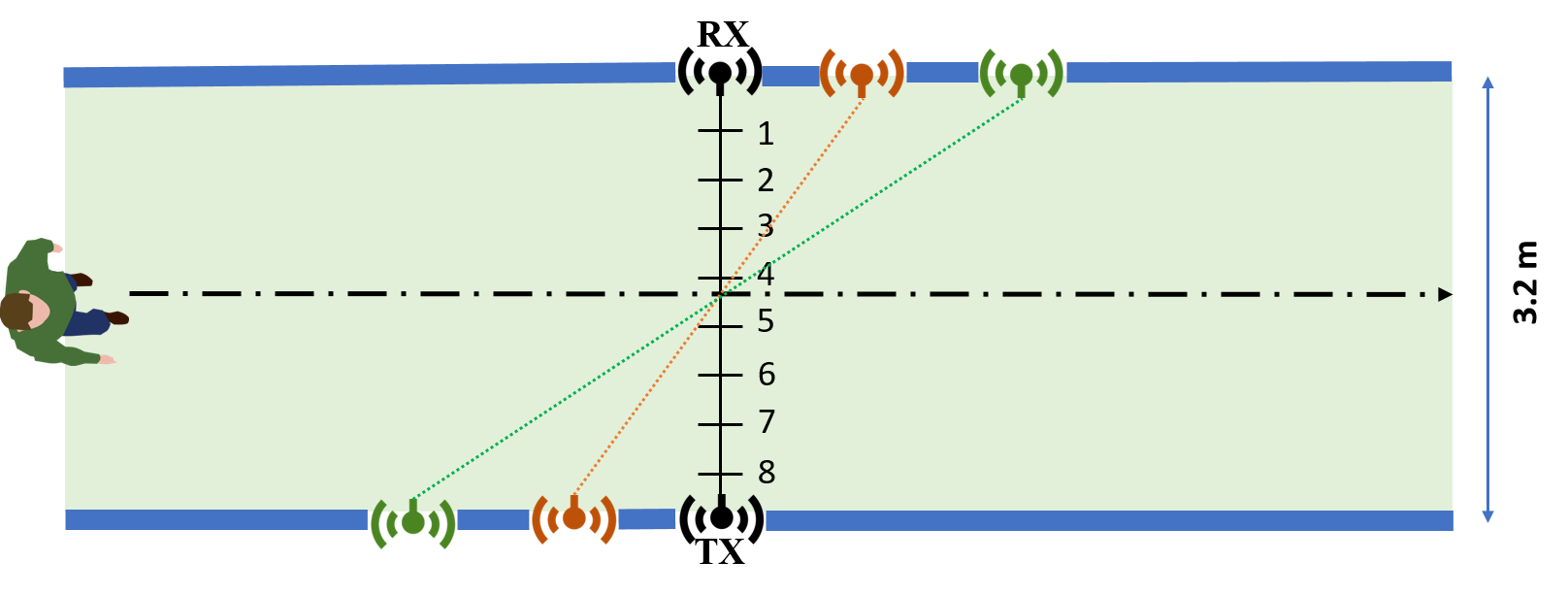}
  \caption{Experimental setup for line crossing.}
  \label{fig:line_setup}
  \end{center}
\end{figure}

We perform experiments in two different indoor hallways of width 3.2~meters, with the nodes kept on either side of the hall way at 1~m high from the ground, as shown in Fig.\ \ref{fig:line_setup}. The single user walks across the link line at a given speed. To keep the subject at a constant speed, we use a smart phone application called  \emph{Metronome} which  beeps periodically to dictate the pace of the subject. We mark the floor along the pathway at a spacing of 61~cm. For each line crossing measurement, the subject walks nearly 14~meters. We conduct the experiment while the subject walks in solitude around the hallway.

\subsection{Results}
First, we evaluate human walking in the range of speeds from 0.3~m/s to 1.81~m/s. We determine the best linear fit between actual velocity and $f_{av}$ and find that $\alpha = 0.88$ for hallway 1 and $\alpha=0.72$ for hallway 2.  Clearly, the best linear fit $\alpha$ may vary per link.  

\begin{figure}[tbhp]
\begin{center}
  \includegraphics[width=0.9\columnwidth]{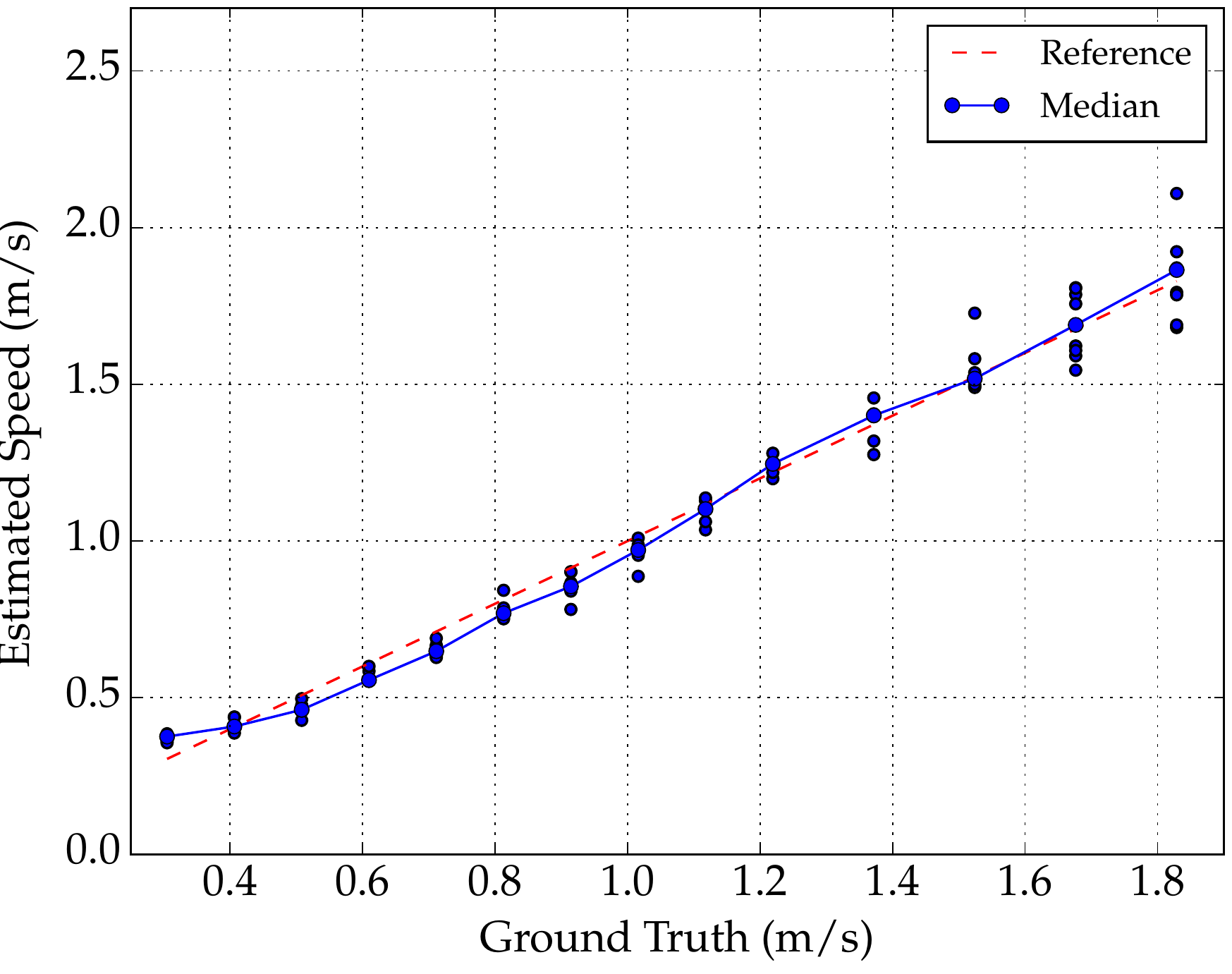}
  \caption{Estimated walking speed vs ground truth when a person crosses a link.}
  \label{fig:speed}
  \end{center}
\end{figure}

Fig.\ \ref{fig:speed} shows the estimated speed $\hat{v}$ as a function of the ground truth walking speed. It is noted that the RMSE is 5.13~cm/s. For a person walking through a hallway at a constant rate, this means that we can accurately estimate a person's distance from the line and be off by less than 1 m for a period of 20 seconds.

\begin{figure}[tbhp]
\begin{center}
  \includegraphics[width=0.9\columnwidth]{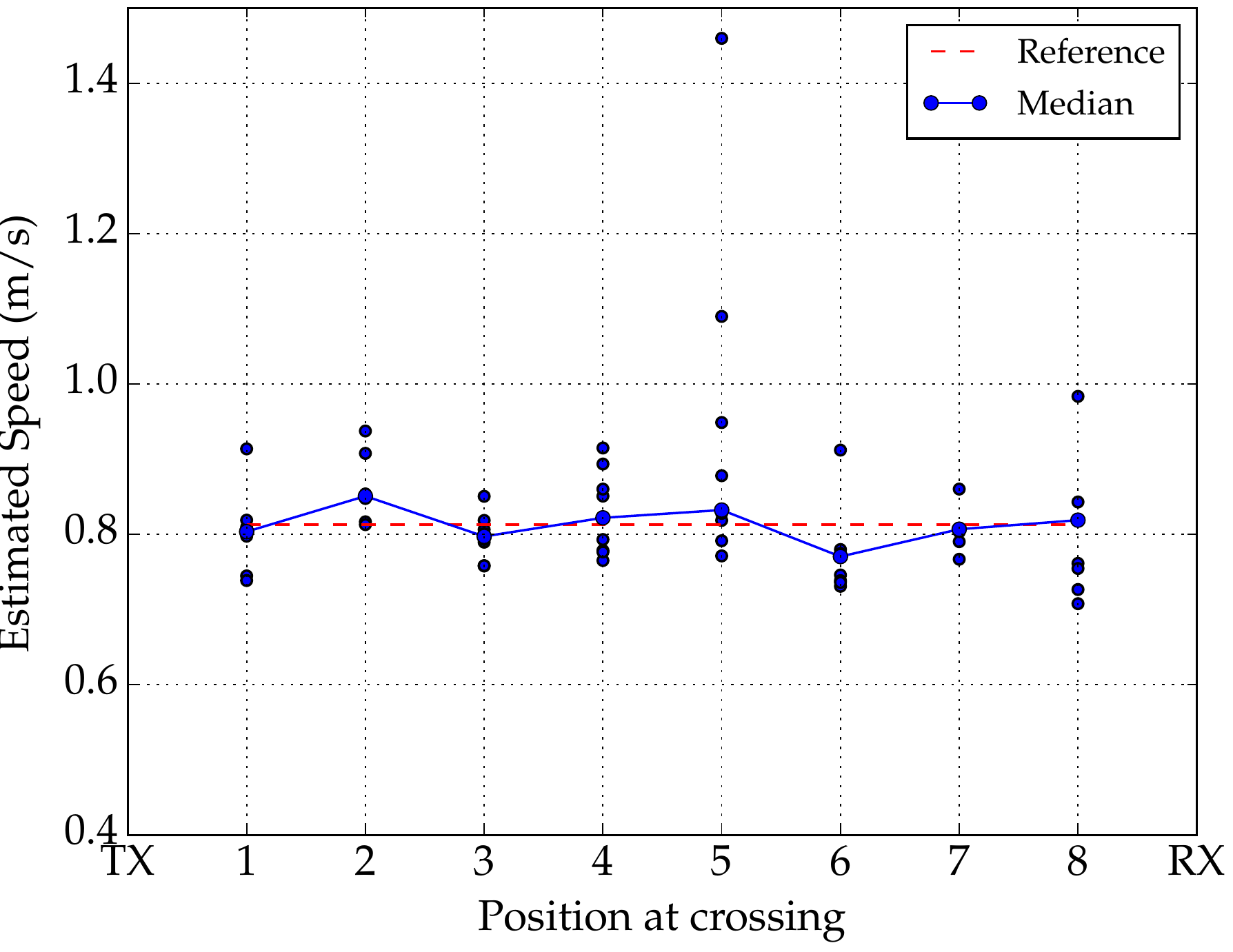}
  \caption{Estimated walking speed vs.\ ground truth when a person crosses a link at different position along the link line.}
  \label{fig:pos}
  \end{center}
\end{figure}

We also show the estimated speed for a person crossing the link line at different places, marked as 1 through 8 in Fig.\ \ref{fig:line_setup}, at a constant speed of 0.813~m/s. As shown in Fig.\ \ref{fig:pos}, the median of the estimated speed is within 0.05~m/s from the correct value.  The physical model of  \cite{rampa2015physical} provides analytical justification for this result.  In that model, 
assuming a transmitter and a receiver are located at ($0, 0$) and ($0, d$), for a person walking at a constant speed, the average RSS frequency at the link line crossing at $y$ meters away from the transmitter is directly proportional to $\frac{1}{y(d-y)}$. This implies that the average RSS frequency is nearly constant with crossing position $y$  unless the person is very close to either transceiver. 

Next, we show that the value of $\hat{v}$ is not a strong function of the angle between the person's walking path and the line between the transmitter and receiver. Fig.\ \ref{fig:line_setup} shows alternative locations for the transmitter and receiver such that the angle between the link line and the person's path is not 90$^o$. We define path angle as the angle between the link line and person's path. We perform multiple line crossing experiments with varying path angles and a person walking 0.813 m/s.  

\begin{figure}[tbhp]
\begin{center}
  \includegraphics[width=\columnwidth]{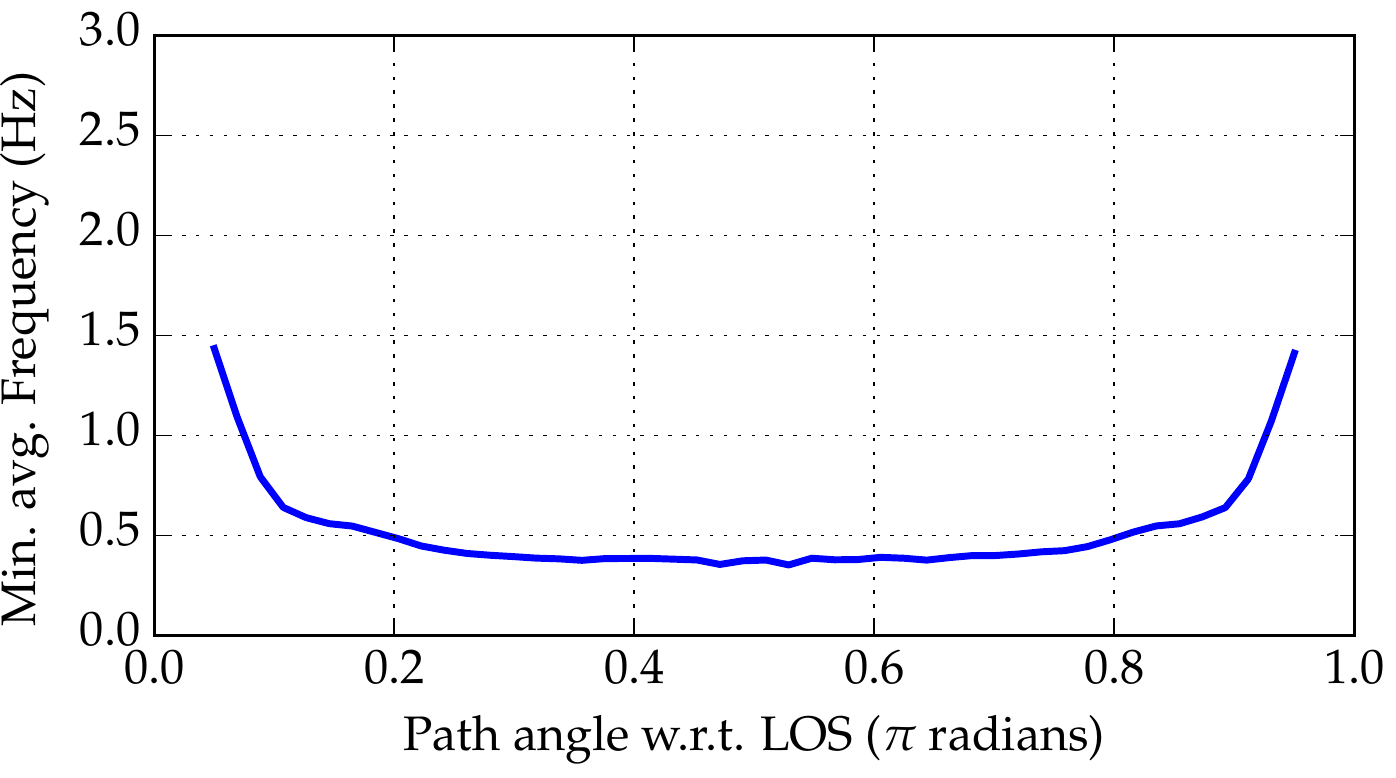}
  \caption{Simulated average frequency vs.\ path angle with speed  0.813~m/s}
  \label{fig:ang_freq}
  \end{center}
\end{figure}

\begin{figure}[tbhp]
\begin{center}
  \includegraphics[width=\columnwidth]{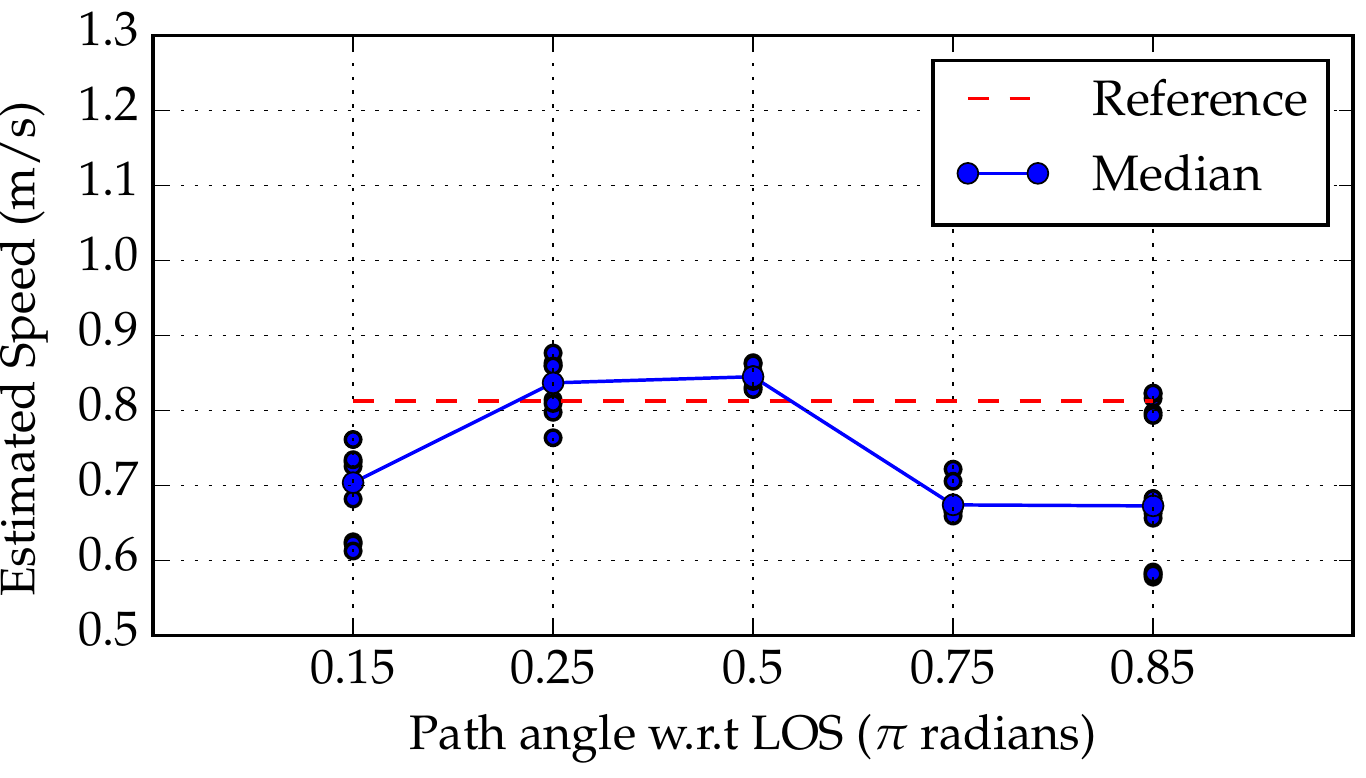}
  \caption{Estimated speed vs.\ path angle from RSS measured for a person walking with a speed of 0.813 m/s.}
  \label{fig:ang_measured}
  \end{center}
\end{figure}

Fig.\ \ref{fig:ang_freq}  shows simulated average RSS frequency at time of crossing as a function of path angle. We note that the value of $\hat{v}$ as a person crosses a link changes only slightly for most path angles except paths close in angle to the link line. For small path angle, the average frequency at the time of crossing becomes high. This is primarily because paths parallel to the link line result in only small changes in RSS, and noise in the RSS accounts for higher average frequency. Fig.\ \ref{fig:ang_measured} shows speed estimated from RSS measured when a person crosses a link line at a speed of 0.813~m/s for different path angles. The results show that there is an RMSE of 10.6~cm/s as a result of varying path angles. The difference between simulated and measured results may arise from the change in person's shadow shape when the path angle changes.
The results show that the average frequency of RSS signal measured when a person crosses a link line allows accurate walking speed estimation despite changes in path angle or crossing position with respect to the transceivers.

\section{Discussion}
\label{sec:discussion}

In the past, RSS measurements from standard radio transceivers have been utilized in device-free localization and breathing monitoring. However, previous work has not been able to determine heart rate, classify gestures accurately, or estimate human speed from single-channel RSS measurements on standard radio transceivers. Part of the problem is that standard RSSI values are quantized to 1 dB, which is too large of a quantization bin for many applications \cite{luong2016rss}.  Our RF sensing system does use fine-grained RSS measurements, and does so using a narrowband signal and a low transmit power.  This paper shows that these applications that are typically seen as requiring multichannel and MIMO measurements, fundamentally don't required such high bandwidth and high dimensionality of radio channel measurements.  One can achieve high performance with limited bandwidth.

However, the use of single-channel RSS measurements for these applications brings the following limitations.   

\textbf{Multiple Users}: Our RF sensing system is shown to determine heart rate, detect gestures, or estimate speed only for a single user. Nearby motions and/or interfering users affect the performance of RF sensing. VitalRadio \cite{adib2015smart} enables vital sign monitoring for multiple users. However, it relies on the use of FMCW radar to separate the reflections arriving from different users into different buckets based on the distance between the users and the device.   Pu \emph{et al.} \cite{pu2013whole} uses MIMO to measure a high dimensional signal which is then able to identify gestures when multiple people are present.  The ability of a single-link narrowband RSS-based system to be able to separate users has not been tested, and we would believe that it would be a very challenging problem. However, we note that our system's low bandwidth enables a higher density of (FDMA) links which would enable ubiquitous, multi-person operation, whereas use of many WiFi or FMCW channels would be more problematic for spectral availability.

\textbf{Non-stationary Users.} Our RF sensing system detects heartbeat for only for stationary users. Since heartbeat estimation involves detecting mm-level movements, the changes in RSS due to a person's pulse are overwhelmed by larger body motions. 

\textbf{Offline Calibration.} To estimate the speed of a subject using our RF sensing system, we currently require calibration for each link. Future work might explore the relationship of the parameter with other measurable link characteristics, or auto-calibration methods. For gesture recognition, change in the path length may alter multipath channel characteristics, and, thus, the features for particular gestures.  Research is required to find more link-independent features and to address the training and re-training requirements of the system.

\textbf{Sensing Range.} Heart rate monitoring using our system requires the subject to be within at most 1~m away from the closest transceiver to keep the signal power high enough to be able to accurately track pulse.  Future work should increase the center frequency so that the antenna size can be reduced and also so that the mm-level vibrations of the skin are larger fractions of the wavelength.  

\textbf{Power and Cost.}
We show that data from the TI CC1200, a low-power, low-cost transceiver IC, is capable of enabling several RF sensing applications.
However, our implementation uses a Beaglebone as the processor. While it is a functional prototype and a complete system, it is not cost, power and size efficient. For the future, the Beaglebone should be replaced by a low-cost real-time microprocessor.  We believe the contributions in this paper could be implemented on a Cortex-M4 with Floating Point Unit (FPU) support, which would cost about 20 EUR per unit, would be smaller than the Beaglebone, and would easily operate with a single cell battery.

We anticipate that, through future work, many of the limitations of the proposed system can be overcome.

\section{Conclusion}
\label{sec:conclusion}
In this paper, we present an RF sensing system using two low-cost single carrier radio transceivers which operate using 11.26 kHz of RF bandwidth. We show that RSS measurements captured from these devices enable accurate heartbeat detection, gesture recognition, and human speed estimation, despite their efficient spectrum utilization.  We develop a method to estimate a person's heart rate from single-channel RSS measurements.  We test a RSS-based gesture recognition which exploits  DWT analysis to extract features, and show that it attains an average accuracy of  85\% across three subjects.  We also demonstrate that human walking speed can be estimated with an error of 5~cm/s, using a novel algorithm which operates on the spectrogram of the recorded RSS data. Our low-cost RF sensing system performs nearly as well as reported results from several state-of-the-art systems while using three orders of magnitude less bandwidth.  Without reducing the very high bandwidth utilization of the state-of-the-art RF sensing systems, we believe that it will be very difficult to realize ubiquitous RF sensing as envisioned by many in the research area.  We believe our results provide an important proof-of-concept to show that low-cost and low-bandwidth sensing is possible and may be an enabler for ubiquitous RF sensing.
\section*{Acknowledgement}
This material is based upon work supported by the US National Science Foundation under Grant Nos. 1407949, 1622741 and 1564287.

\balance
\bibliographystyle{ACM-Reference-Format}
\bibliography{sigproc.bib}

\end{document}